%% file: paper.tex
\documentclass[aps,prd,preprint,showpacs,groupedaddress,superscriptaddress,nofootinbib]{revtex4}

\pdfoutput=1
\usepackage{graphicx}  
\usepackage{amsmath}

\input{macros.tex}

\begin{document}
\title{Probing the chiral regime of $N_f=2$ QCD with mixed actions}

\author{F.~Bernardoni}

\affiliation{Instituto de F\'{\i}sica Corpuscular, CSIC-Universitat de Val\`encia\\
		Apartado de Correos 22085, E-46071 Valencia, Spain}
\affiliation{NIC, DESY, Platanenallee 6, 15738 Zeuthen, Germany}

\author{N.~Garron}
\affiliation{School of Physics and Astronomy, University of Edinburgh\\
Edinburgh EH9 3JZ, United Kingdom}

\author{P.~Hern\'andez}
\affiliation{Instituto de F\'{\i}sica Corpuscular, CSIC-Universitat de Val\`encia\\
		Apartado de Correos 22085, E-46071 Valencia, Spain}

\author{S.~Necco}
\affiliation{CERN, Physics Department, TH Division\\
        CH-1211 Geneva 23, Switzerland}

\author{C.~Pena}
\affiliation{Dpto. de F\'{\i}sica Te\'orica and Instituto de F\'{\i}sica Te\'orica UAM/CSIC\\
        Universidad Aut\'onoma de Madrid, Cantoblanco E-28049 Madrid, Spain}





\begin{abstract}
We report on our first experiences with a mixed action setup with overlap valence 
quarks and non-perturbatively $\Oa$ improved Wilson sea quarks. For the latter
we employ CLS $N_f=2$ configurations with light sea quark masses at small lattice 
spacings. Exact chiral symmetry allows to consider very light valence quarks and
explore the matching to (partially quenched) Chiral Perturbation Theory (ChPT) in a mixed
$\epsilon$/$p$-regime. We compute the topological susceptibility and the low-lying spectrum of the massless Neuberger-Dirac operator 
for three values of the sea quark mass, and compare the sea quark mass dependence to NLO ChPT in the mixed regime. 
This provides two different determinations of the chiral condensate,
as well as information about some NLO low-energy couplings. Our results allow to
test the consistency of the mixed-regime approach to ChPT, as well as of the mixed action 
framework.
\end{abstract}

\maketitle

\input{intro.tex}
\input{theory.tex}

\input{num.tex}

\input{fits.tex}
\input{concl.tex}

\cleardoublepage
\appendix
\input{app1.tex}

\input{app2.tex}

\cleardoublepage
\input{figtab.tex}

\cleardoublepage

\bibliographystyle{apsrev}

\bibliography{biblio}

\end{document}

%% file: macros.tex
\newcommand{\ba}{\begin{array}}
\newcommand{\ea}{\end{array}}

\newcommand{\refig}[1]{Fig.~\ref{#1}}

\newcommand{\dif}{{\rm d}}

\newcommand{\Dslash}{\relax{\kern+.25em / \kern-.70em D}}

\newcommand{\fm}{{\rm fm}}

\newcommand{\GeV}{{\rm GeV}}

\newcommand{\Real}{\relax{\mathsf{\Gamma\kern-.35em R}}}
\newcommand{\Int}{\relax{\mathsf{Z\kern-.40em Z}}}

\newcommand{\gbar}{\kern1pt\overline{\kern-1pt g\kern-0pt}\kern1pt}
\newcommand{\mbar}{\kern2pt\overline{\kern-1pt m\kern-1pt}\kern1pt}
\newcommand{\obar}[1]{\kern3pt\overline{\kern-2pt #1\kern-0pt}\kern1pt}

\newcommand{\Oa}{\mbox{O}(a)}

\newcommand{\abar}{\kern1pt\overline{\kern-1pt a\kern-0.5pt}\kern1pt}



%% file: intro.tex
\section{Introduction}

Thanks to the theoretical and algorithmic improvements of recent years,
and to the ever increasing computational power available,
state-of-the-art Lattice QCD simulations now easily reach 
dynamical pion masses in the 200--300~MeV ballpark.\footnote{See e.g. the review~\cite{Jung:2010jt} at Lattice 2009.} In this mass region 
the effective description of the dynamics of pseudo-Goldstone bosons
at low energies by means of chiral perturbation theory (ChPT) is expected to work
well at a quantitative level. This gives rise to a fertile interaction: by matching
Lattice QCD and ChPT results it is possible, on the one hand, to test the effective
description vs. the fundamental theory; and, on the other hand,
low-energy constants (LECs) can be determined from first principles, thus providing
a sounder foundation to phenomenological applications of ChPT.

One particularly interesting aspect of the matching between QCD and ChPT is
the role of finite volume effects~\cite{Gasser:1986vb,Gasser:1987ah,Neuberger:1987zz,Neuberger:1987fd}
(we will always assume that the theory lives in
an Euclidean four-volume $V=L^3 \times T$). While for large enough values of
$L$ (one typical estimate is $M_\pi L \gtrsim 4$) the latter are expected
to be strongly suppressed, and give rise only to small corrections to the infinite
volume expansion in powers of pion momenta, the situation changes completely when the 
Compton wavelength of pions approaches $L$, i.e. $M_\pi L \sim 1$. In this
regime slow pion modes, strongly affected by the finite volume, dominate the path
integral in the effective theory, and the expansion in powers of $M_\pi^2/\Lambda_\chi^2$
breaks down. A new power-counting for this $\epsilon$-regime was proposed in~\cite{Gasser:1987ah}, which implies a
rearrangement of the chiral expansion, in which quark mass effects are suppressed relative to 
volume effects.  As a consequence, less LECs appear at any given order in 
the expansion relative to the infinite volume case, and the approach to the chiral limit is therefore more universal.
 This regime leads to a very
different setup for the determination of LECs, which offers both the potential
to obtain cleaner computations of some of the latter (those whose effects are unsuppressed in the quark mass),
and a cross-check of the systematic uncertainties of ``infinite'' volume studies.

Another key property of finite volume chiral regimes ($\epsilon$-regime) is that the partition function of ChPT (at leading-order in the $\epsilon$-expansion) has been shown to be equivalent to a random matrix theory (RMT) \cite{Shuryak:1992pi,Verbaarschot:1993pm,Verbaarschot:1994qf,Verbaarschot:2000dy}, where many analytical predictions can be obtained for spectral quantities, such the spectral density  or  the distribution of individual eigenvalues. These predictions are expected to be valid also for the spectrum of the Dirac operator in this regime, and 
have  been tested both in quenched~\cite{Edwards:1999ra,Bietenholz:2003mi,Giusti:2003gf}, $N_f=2$ \cite{DeGrand:2006nv,Lang:2006ab,Fukaya:2007yv} and $N_f=2+1$  \cite{Hasenfratz:2007yj,Fukaya:2009fh} QCD. Since RMT predictions depend on one free parameter that corresponds to the chiral condensate (at the leading-order in the matching of ChPT and RMT), they provide yet another way
of studying chiral symmetry breaking, using simple spectral observables.

As we will see the matching of ChPT in the $\epsilon$-regime holds up to NLO in the chiral expansion, both for full and partially-quenched (PQ) situations as long as some quarks remain in the $\epsilon$-regime, because the path integral is dominated by the zero-modes of the {\it lighter} pions. The matching of QCD to RMT can therefore be extended to unphysical situations which are however more favorable from the computational point of view. 

Obviously enough, an adequate treatment of chiral symmetry on the lattice 
is especially relevant in this context. While simulations of $N_f=2(+1)$ QCD with
full chiral symmetry have proven feasible, they are still limited to relatively small
values of the inverse lattice spacing and/or physical volume~\cite{Fukaya:2010ih}.
A way to overcome this is
to use a mixed action approach~\cite{Bar:2002nr,Bar:2003mh}, in which chiral symmetry is exactly preserved at the
level of valence quarks only. Our aim is to develop such a framework by considering
overlap valence quarks on top of $N_f=2$ Coordinated Lattice Simulations (CLS) ensembles,\footnote{https://twiki.cern.ch/twiki/bin/view/CLS/WebHome} obtained from simulations with non-perturbatively $\Oa$ improved Wilson sea quarks.

We will use this method to study the matching of QCD to ChPT and RMT in a mixed regime, 
in which sea quark masses are in the $p$-regime and valence quark masses are in the $\epsilon$-regime~\cite{Bernardoni:2007hi,Bernardoni:2008ei}. From this matching we will be able to extract the $N_f=2$ low-energy couplings $\Sigma$ and $L_6$. 

Furthermore, the use of overlap fermions allows us to measure the topological susceptibility of the dynamical Wilson configurations. The dependence of this quantity on the sea quark mass has been derived in ChPT at NLO in infinite volume in \cite{Chiu:2008jq}, and has been shown to depend on $\Sigma$ and on 
 a combination of $L_6$, $L_7$ and $L_8$. We will show how this prediction can be easily obtained in the mixed regime of ChPT and we will test it from the measured distribution of the topological charge. 

Obviously, mixed actions also have huge potential for phenomenological applications in which the
exact preservation of chiral symmetry is greatly advantageous, e.g. to simplify
the renormalisation of composite operators entering hadronic weak matrix elements.
Along this line, first data for standard two- and three-point functions, as well as
for correlation functions computed in the chiral limit via saturation with
topological zero modes~\cite{Giusti:2003iq,Hernandez:2008ft}, will be covered in upcoming publications.

The structure of the paper is as follows. In Section~\ref{sec:chpt} we review the main results from \cite{Bernardoni:2007hi,Bernardoni:2008ei} on the mixed-regime of ChPT and collect the results needed for our work, in which the sea quarks are degenerate and lie in the $p$-regime and  valence quarks are in the $\epsilon$-regime. We will show that, up to NLO in the mixed-regime expansion, the partition functional matches onto a RMT, where the free parameter depends on the sea quark mass in a way that can be predicted from the matching. In Section ~\ref{sec:num} we present our numerical results for the low-lying spectrum of the overlap operator and the topological susceptibility on 
Wilson sea quarks and compare with the predictions of RMT and ChPT. In Section ~\ref{sec:fits} we present our results for the fits to ChPT predictions and extract the low-energy couplings. 

%% file: theory.tex
\section{Probing the deep chiral regime with mixed actions}
\label{sec:chpt}

\subsection{Mixed chiral regimes}

While the exploration of chiral finite volume regimes ideally involves simulations with 
extremely light sea quarks, it is still possible to access them in a situation in
which sea quarks have moderately larger masses. 

The first step is to formulate ChPT
in a so-called mixed regime~\cite{Bernardoni:2007hi,Bernardoni:2008ei}, in which $N_s$ quarks  have masses such that the
$p$-regime requirement $m_s\Sigma V\gg 1$ is satisfied, while $N_l$ quarks  have masses
that fulfill the $\epsilon$-regime condition $m_l\Sigma V \lesssim 1$.  An appropriate power counting for this regime was first introduced in \cite{Bernardoni:2007hi,Bernardoni:2008ei}:
\begin{eqnarray}
\label{eq:counting1}
p_\mu \sim {\cal O}(\epsilon),\;\;\;
L,T \sim {\cal O}(1/\epsilon),\;\;\;
m_{l}  \sim {\cal O}(\epsilon^4),\;\;\;
m_s  \sim {\cal O}(\epsilon^2).
\label{eq:pc}
\end{eqnarray}
The partition function and meson correlators were computed to next-to-leading (NLO) order
according to this power-counting, both in the context of fully dynamical quarks and also in various partially-quenched situations.
The relevant PQ zero-mode integrals  where also studied in \cite{Damgaard:2007ep}. 
In this work, we want to keep the sea quarks in the $p$-regime and the valence quarks in the $\epsilon$-regime.  
We briefly describe the results for this situation, and we refer to the original papers for details on the computations. 
  
  As is common in the $\epsilon$-regime of PQChPT, we need to consider sectors of fixed topological charge. We will be dealing therefore with the replica method in which 
  the PQ limit is obtained in the limit $N_l \rightarrow 0$. 
  
  The starting point is the parametrization of the Goldstone manifold according to the power-counting above. It can be shown that the  
 non-perturbative  zero-modes can be parametrized by a constant matrix, $\overline{U}_0 \in U(N_l)$ that together with various perturbative modes, $\xi(x), \bar{\eta} \sim {\mathcal O}(\epsilon)$ span the full Goldstone manifold ($\xi$ contains all the non-zero momentum modes, while $\bar{\eta}$ 
parametrizes the only perturbative zero-mode --- for details see \cite{Bernardoni:2007hi,Bernardoni:2008ei}). 
  The partition function  in a sector of charge $\nu$ is given by  \cite{Bernardoni:2007hi,Bernardoni:2008ei}
    \begin{eqnarray}
Z_{\nu}\simeq \int 
\left[d \xi\right] \left[d \bar{\eta}\right] \int_{U(N_l)} \left[d\overline{U}_0\right] ~J(\xi)~ 
 \det (\overline{U}_0)^{\nu} Ê\exp\left(-\int d^4 x {\mathcal L}(\xi,\bar{\eta},\overline{U}_0) \right). \nonumber\\
 \end{eqnarray}
 $J(\xi)$ is the Jacobian associated with the parametrization \cite{Hansen:1990un,Bernardoni:2007hi,Bernardoni:2008ei}, and both the Lagrangian and the Jacobian can be perturbatively expanded in powers of $\xi$ and $\bar{\eta}$. 
 The Lagrangian has an expansion in $\epsilon$ of the form:
 \begin{eqnarray}
 {\mathcal L} &=&  \mathcal{L}^{(4)} +  \mathcal{L}^{(6)} + ...,
 \end{eqnarray}
 with terms up to $\mathcal{O}(\epsilon^4)$, up to $\mathcal{O}(\epsilon^6)$, etc , while $J(\xi) = 1 +{\mathcal O}(\epsilon^2)$.

The leading-order Lagrangian is found to be:
 \begin{eqnarray}
  \mathcal{L}^{(4)} &\equiv&  {\rm Tr}\left[\partial_{\mu}\xi \partial_{\mu}\xi\right] -\frac{\Sigma}{2}{\rm Tr} 
\left[ P_l \mathcal{M} P_l (\overline{U}_0+\overline{U}_0^{\dagger})\right]   
+  {2 \Sigma \over F^2} {\rm Tr}\left[  P_s \mathcal{M} P_s  \left(\xi - {F \over 2}{\bar{\eta} \over N_h} P_s\right)^2\right] + i {\nu \over V} \bar{\eta} , \nonumber\\\label{eq:l4}
 \end{eqnarray}
where $P_{l,s}$ are the projectors on the light and heavy quark sectors and $\mathcal{M}$ is the quark mass matrix. At this order, there is  a factorization of the perturbative and non-perturbative modes.  

An important observation from eq.~(\ref{eq:l4}) is that the dependence on the non-perturbative modes, $\overline{U}_0$, and on the light quark mass is identical to that of a theory with $N_l$ quarks in the $\epsilon$-regime, but with the low-energy coupling, $\Sigma$, corresponding to a theory with $N_f=N_l+N_s$ flavours. This is to be expected since the heavier $p$-regime quarks behave, at the lowest energies, as decoupling particles that can be integrated out, but they are not quenched \cite{Bernardoni:2007hi}.  

The quadratic form of the perturbative modes justifies their scaling with $\epsilon$. The last term in eq.~(\ref{eq:l4}) could be treated as a perturbation as long as $\nu \sim \mathcal{O}(\epsilon^0)$. However, in the partially-quenched case $N_l = 0$ that we will be considering here, the 
  distribution of topological charge  is, on average, controlled by the sea quarks only.  Indeed the $\nu$ dependence of the leading-order function is found to be (after integrating over the perturbative modes)
  \begin{eqnarray}
  Z^{LO}_\nu \propto \exp\left( -{\nu^2\over V F^2 } \sum_s { 1 \over M_{ss}^2}\right) \int_{U(N_l)} \left[d\overline{U}_0\right]   \det (\overline{U}_0)^{\nu}  \exp \left( \frac{\Sigma}{2}{\rm{Tr}}\left[ P_l \mathcal{M} P_l (\overline{U}_0+\overline{U}_0^{\dagger})\right] \right), \nonumber\\
  \label{eq:zlonu}
  \end{eqnarray}
  where 
  \begin{eqnarray}
M_{ss}^2 \equiv {2 m_s \Sigma \over F^2}.
\end{eqnarray}
In the case $N_l \rightarrow 0$, the integral over the zero-modes is exactly one (as in the quenched case) therefore all the $\nu$ dependence of the LO partition functional is in the explicit  gaussian factor in eq.~(\ref{eq:zlonu}). The Leutwyler-Smilga result \cite{ls} for the topological susceptibility is obtained:
  \begin{eqnarray}
  \langle \nu^2 \rangle = {1 \over 2} V F^2 {1 \over \sum_s {1 \over M_{ss}^2}} \sim \epsilon^{-2},
  \end{eqnarray}
 a scaling that makes the last term in eq.~(\ref{eq:l4}) of $\mathcal{O}(\epsilon^4)$, and therefore of leading order.  

In fact we can easily push the computation of the topological susceptibility to NLO. In \cite{Mao:2009sy,Aoki:2009mx}, the topological susceptibility has  been computed in ChPT to NLO when all the quarks are in the $p$-regime. The same result  should be obtained in the mixed-regime case when $N_l \rightarrow 0$, since only the dynamical $p$-regime quarks can contribute to the distribution of topological charges when $N_l =0$.  Indeed the NLO partition functional of ChPT in the mixed regime at fixed topology can be computed straightforwardly according to the power-counting rules of the mixed-regime, and the $\nu$ dependence can be explicitly determined. It turns out that for $N_l \rightarrow 0$, all the $\nu$-dependence comes from the integration of the perturbative modes $\xi$ and $\bar{\eta}$.   The result for $N_s$ degenerate quarks is found to be
\begin{eqnarray}
\langle \nu^2\rangle = {m_s \Sigma V\over N_s} & &\left[ 1-  {N_s^2 -1 \over N_s} \left({M_{ss}^2\over 16 \pi^2 F^2} \log\left( {M_{ss}^2 \over \mu^2}\right) + g_1(M_{ss},L,T)\right) \right.\nonumber\\
&+& \left. {16 M_{ss}^2 \over F^2} \left( L^r_8(\mu)+ N_s L^r_6(\mu) + N_s L^r_7(\mu)  \right) \right],
\label{eq:nu2nlo}
\end{eqnarray}
 where the function $g_1(M, L, T)$ contains the finite volume corrections to the closed pion propagator; its explicit definition can be found in \cite{Hasenfratz:1989pk}.
This result agrees with the NLO results of  \cite{Mao:2009sy, Aoki:2009mx}. Note the appearance of  $L_7^r(\mu)$,  for which no prediction has  yet been obtained on the lattice. 

\vspace{0.5cm}

In \cite{Shuryak:1992pi,Verbaarschot:1993pm,Verbaarschot:1994qf,Verbaarschot:2000dy}, it was shown that the $\epsilon$-regime zero-mode partition function at LO
 is that of a Random Matrix Theory (RMT) of matrices of size $N$, that depends only on the number of flavours, $N_l$, and the corresponding mass parameters $\hat{m}_l$, viz.
 \begin{eqnarray}
  \int_{U(N_l)} \left[d U\right]   \det (U)^{\nu}  \exp\left( \frac{\Sigma}{2}{\rm Tr} \left[ \mathcal{M} (U+U^{\dagger})\right] \right)  \simeq {\rm RMT}_N\{N_l,\hat{m}_l\},
  \label{eq:shuryak}
 \end{eqnarray}
where ${\mathcal M} = \delta_{ij} m_i$ is the $N_l \times N_l$ mass matrix and there must be an identification $N \hat{m}_i = m_i \Sigma V$.

 From this relation, the microscopic spectral density  of the Dirac operator, as well as higher order spectral correlation functions, can be related to those quantities computed  in the corresponding RMT. Furthermore, the distribution of individual low-lying eigenvalues of the massless Dirac operator can also be predicted from this equivalence \cite{Nishigaki:1998is,Damgaard:2000ah,Akemann:2008va}, providing an efficient method to determine the chiral condensate, $\Sigma$. This relation has been tested in the quenched approximation and a good agreement has been found for volumes above $1.5$ fm or so \cite{Giusti:2003gf}. In dynamical simulations, it has also been tested in \cite{DeGrand:2006nv,Lang:2006ab,Fukaya:2007yv} for $N_f=2$ and in \cite{Hasenfratz:2007yj,Fukaya:2009fh} for $N_f=2+1$.
More details on the RMT formulation will be given in the Section \ref{sec:rmt}.

\begin{figure}[t!]
\begin{center}
\includegraphics[width=10cm,angle=-90]{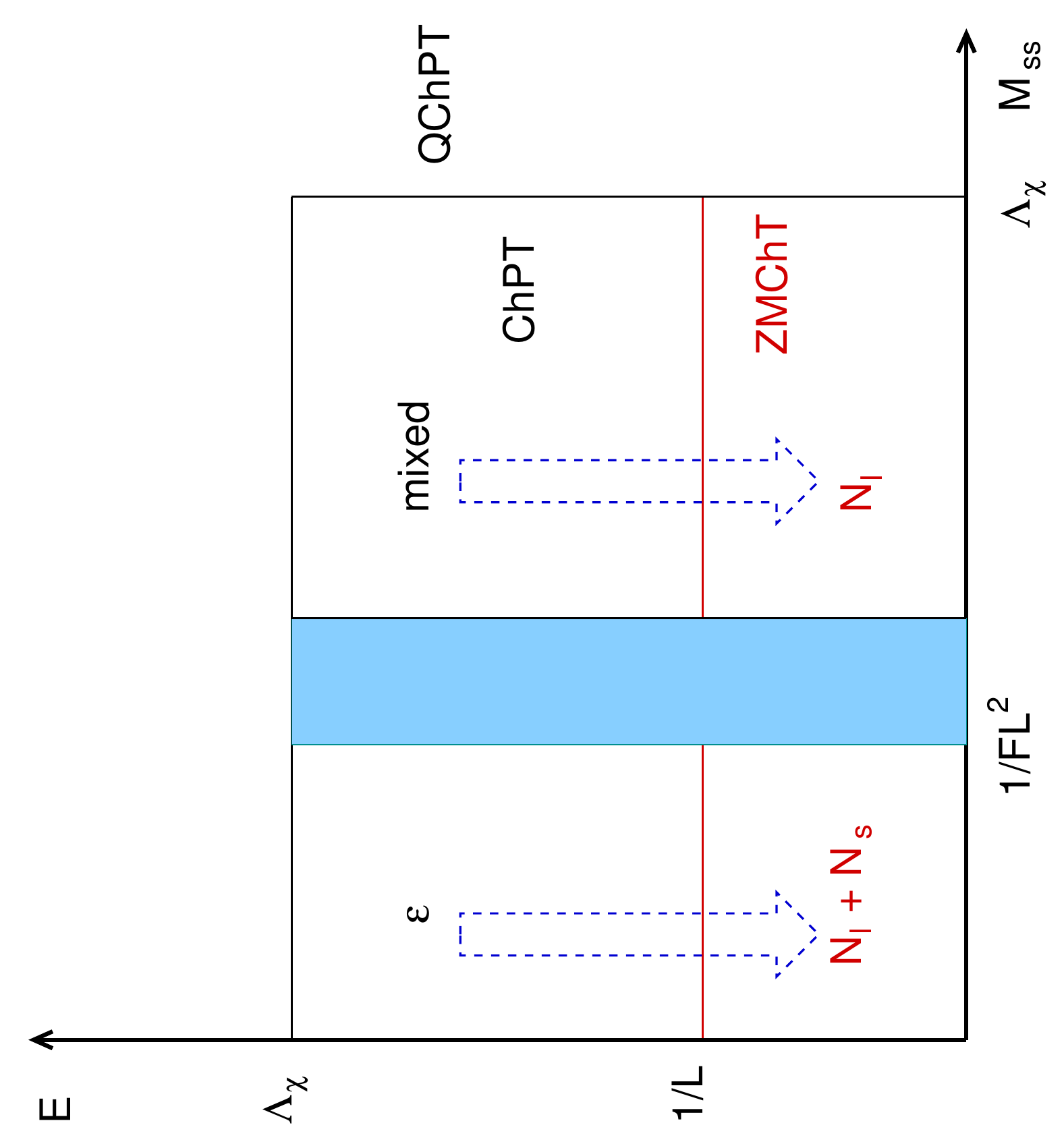}
\caption{Chiral regimes of QCD showing the range of validity of the zero-mode chiral theory (ZMChT), which is equivalent to a RMT, and is obtained from ChPT by integrating out the non-zero momentum modes. $\Lambda_\chi\simeq 4\pi F$ represents the chiral symmetry breaking scale.
The LECs of the ZMChT ($\Sigma_{\rm eff}$) can be derived from the LECs of ChPT from matching, which can be done in two regimes depending on the size of the sea quark mass. For $M_{ss} \ll 1/F L^2$, the $\epsilon$ regime is appropriate, while for $M_{ss} \geq 1/F L^2$  the mixed-regime has to be considered. The resulting ZMChTs have different number of flavours in the two cases.}. 
\label{fig:zmcht}
\end{center}
\end{figure}

\vspace{0.5cm}
The rationale for expecting a matching of QCD to RMT relies  on the existence of a  regime where the chiral effective theory  simplifies to a theory containing only the Goldstone zero-modes, as depicted in Fig.~\ref{fig:zmcht}. In fact,  if we consider ChPT in the usual $\epsilon$-regime or in the mixed-regime above, there is a hierarchy of scales $M_{vv} \ll L^{-1}$, which implies that we can integrate out the heavy scale $L^{-1}$  to obtain a theory of zero-modes only, which we could call ZMChT (zero-mode chiral theory). We can obtain this theory from the full ChPT integrating the heavy modes order by order in the
$\epsilon$-expansion. The difference between doing this matching in the $\epsilon$  or the mixed-regime is the different assumption on the scaling of 
$M_{ss}$. In the former case, $M_{ss} \ll L^{-1}$ and this scale is not integrated out (the ZMChT has therefore  $N_l+N_s$ flavours), while in the latter  $L^{-1} \sim M_{ss}$ and the zero-modes of the sea pions must be integrated out as well (the ZMChT has then $N_l$ flavours). 

  The matching of ChPT and ZMChT at LO in the mixed-regime can be easily derived from eq.~(\ref{eq:l4}): the ZMChT  is simply ChPT at this order without the heavy modes (the integration over them gives an irrelevant normalization factor):
\begin{eqnarray}
\left. Z_\nu^{ZMChT} \right|_{LO} \propto \int_{U(N_l)} [d \overline{U}_0] (\det \overline{U}_0 )^\nu \exp\left({\Sigma V\over 2} {\rm Tr}\left[ P_l {\mathcal M} P_l \left(\overline{U}_0 + \bar{U}^\dagger_0\right)\right]\right).
\label{eq:zmchtlo}
\end{eqnarray} 
According to the eq.~(\ref{eq:shuryak}), this partition function is then equivalent to an $N_l$  RMT. In particular, it is important to stress that  ZMChT has $N_l$ flavours, while the full ChPT from which it is derived corresponds to $N_f=N_l+N_s$ flavours.  In particular, for $N_l \rightarrow 0$,  the ZMChT or RMT we expect to find is the {\it quenched} one, while the couplings should be those of an $N_f=N_s$ theory. 

\vspace{0.5cm}

The matching at NLO still does not modify the structure of the ZMChT theory. We could have anticipated this by realizing that there are no operators in the list of Gasser and Leutwyler that depend only on a constant $\overline{U}_0$ at ${\mathcal O}(\epsilon^6)$. This does not mean however that there are no corrections, simply that they can be absorbed in the couplings appearing at LO in eq.~(\ref{eq:zmchtlo}), that is $\Sigma$. Indeed 
at NLO, there are corrections to the zero-mode Lagrangian from the ${\mathcal O}(p^6)$ terms 
\begin{eqnarray}
  \mathcal{L}^{(6)} &=& 
  \ldots+
  {\Sigma \over F^2} {\rm Tr}\left[ P_l\mathcal{M} P_{l}  
\left(\xi^2(x) \overline{U}_0 + {\bar U}_0^\dagger \xi^2(x)\right) \right] \\
& -& 16 {\Sigma L_6\over F^4}  {\rm Tr}\left[ P_s \mathcal{M}_s P_s\right]  
{\rm Tr}\left[ P_{l} \mathcal{M} P_{l}  \left( \overline{U}_0 + 
{\bar U}_0^\dagger \right) \right] + \ldots . \nonumber
   \end{eqnarray}
The integrations over the $\xi, \bar{\eta}$ fields  result in a change  $\Sigma \rightarrow \Sigma_{\rm eff}$~\cite{Bernardoni:2008ei},
\begin{eqnarray}
Z_\nu^{ZMChT} |_{NLO} = \int_{U(N_l)} [d \overline{U}_0] (\det \overline{U}_0 )^\nu \exp\left({\Sigma_{\rm eff} V\over 2} {\rm Tr}\left[ {\mathcal M}_l \left(\overline{U}_0 + \bar{U}^\dagger_0\right)\right]\right),
\end{eqnarray}
with (for degenerate sea quarks)
\begin{equation}
{\Sigma_{\rm eff}\over \Sigma} -1 = {1 \over F^2} \left[16 {L_6} N_s M^2_{ss} - {N_l} \bar{\Delta}(0) - {N_s} \Delta(M^2_{ss}/2) + \left( {1 \over N_l}\bar{\Delta}(0) - {N_s \over N N_l} \bar{\Delta}(M^2_\eta)\right)\right],
\end{equation}
where
\begin{equation}
M_\eta^2 \equiv {N_l \over N_f} M_{ss}^2.
\end{equation}
In dimensional regularization, 
\begin{eqnarray}
\Delta(M^2) = {M^2 \over (4 \pi)^2 } \left( - \lambda + \log {M^2 \over \mu^2} \right) + g_1(M, L, T), \;\;\;\;\; \bar{\Delta}(M^2) = \Delta(M^2) -{1 \over V M^2}, 
\end{eqnarray}
and $\lambda$ contains the expected UV divergence
\begin{eqnarray}
\lambda\equiv {1 \over \epsilon} + \log 4 \pi - \gamma +1 - \log \mu^2, \;\;\;\; \epsilon = 2 -d/2, 
\end{eqnarray}
that gets fully subtracted in the usual $\overline{\rm MS}$ scheme. The small $M$ expansion of $\Delta$ gives
\begin{eqnarray}
  \Delta(M^2) = {1 \over V M^2} +  {M^2 \over (4 \pi)^2 } \left( - \lambda - 2 \log { \mu L} \right)  - \sum_{n=1}^\infty {1\over (n-1)!} \beta_n M^{2 (n-1)} L^{2 (n-2)},
\end{eqnarray}
where $\beta_n$ are the shape coefficients that depend only on the ratio $T/L$ \cite{Hasenfratz:1989pk}. Note that the  $M \rightarrow 0$ limit of $\bar{\Delta}(M^2)$ is well defined.

\vspace{0.5cm}

In summary, up to NLO we have found that the ZMChT is equivalent to a RMT. Furthermore, the matching of this theory with ChPT gives the precise dependence on the sea quark mass of the coupling $\Sigma_{\rm eff}$ which is the only free parameter of the RMT theory. Testing this prediction will be one of the main results of this work. 

At this point it is interesting to discuss the possibility to have a smooth transition within the ZMChT regime between the $N_f=N_l+N_s$ effective theory and the 
$N_l$ theory as the scale $m_s$ is increased. The authors of  \cite{Fukaya:2007yv} have assumed that indeed this is possible and have found some evidence that the eigenvalue ratios
seem to follow the dependence on $m_s$ predicted by the RMT or ZMChT. Such expectation would be justified if the conditions were such that $F L \gg 1$, because in this case  the 
 scale $M_{ss}$ can be neglected in the integration of the non-zero modes, as is done in the $\epsilon$-regime. However this is not true in practice, where $F L \sim 1$, and indeed even though the eigenvalue ratios  in \cite{Fukaya:2007yv} showed roughly the dependence on $m_s$ predicted by the RMT (note that $\Sigma_{\rm eff}$ drops in the ratios), this is certainly not true for the eigenvalues themselves. The mixed-regime matching is the correct procedure to account for the correct $m_s$ dependence of $\Sigma_{\rm eff}$, for large enough $m_s$. If $FL$ is not sufficiently large,  there is no warranty that the transition region (vertical band in Fig.~\ref{fig:zmcht}) can be modeled correctly by RMT. For a recent proposal to get predictions in the intermediate region from a resummation of zero-modes see \cite{Damgaard:2008zs}.

\vspace{0.5cm}
 
The distribution of the lowest lying eigenvalues of the Dirac operator is expected to match the prediction of RMT with $N_l$ flavours and  $\Sigma V\rightarrow \Sigma_{\rm eff} V$. $\Sigma_{\rm eff}$ depends on the low-energy couplings of the $N_f=N_l+N_s$ theory. Note that the only NLO coupling entering is $L_6$. 

\begin{figure}
\begin{center}
\includegraphics[width=9cm]{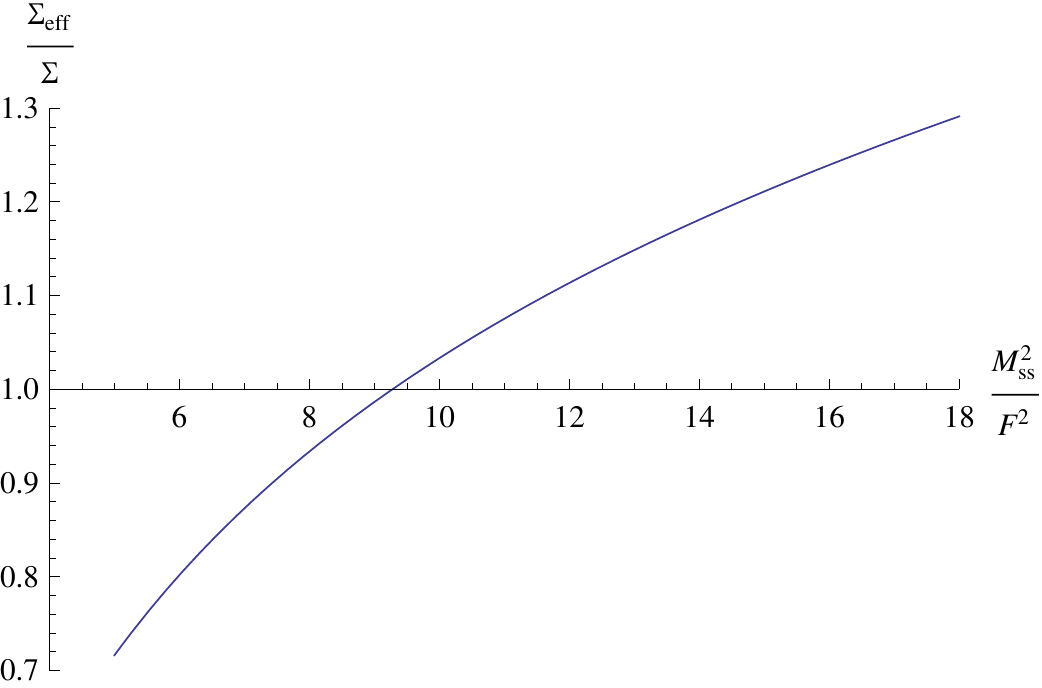}
\caption{$\Sigma_{\rm eff}/\Sigma$ for $N_f=N_s=2$ as a function of $M^2_{ss}/F^2$ for $F=90$ MeV, $L_6^r(M_\rho) = 0.07 \cdot 10^{-3}$.}
\label{fig:sigmaeff}
\end{center}
\end{figure}

Now since we want to consider  the case of quenched light quarks, we have to take the limit $N_l \rightarrow 0$. $\Sigma_{\rm eff}$ has a finite replica limit  given by
\begin{eqnarray}\label{eq:condensate}
\lim_{N_l \rightarrow 0} \Sigma_{\rm eff} &\equiv& \Sigma \,\Bigg\{ 1 + {M_{ss}^2 \over F^2}  \Bigg[ {\beta_2\over N_s} + {\log(\mu V^{1/4}) \over 8 \pi^2 N_s} +
 16 N_s  L_6^r(\mu)-  {N_s \over (4 \pi)^2}  \log \left({M_{ss} \over \sqrt{2} \mu}\right) \Bigg]  \nonumber\\
  && ~~~~~~ -{\beta_1\over N_s F^2 \sqrt{V}} -{N_s\over F^2} g_1\left(M_{ss}/\sqrt{2}, L, T \right)  \Bigg\}.
  \label{eq:sigmaeff}
\end{eqnarray}
 For the case $T/L=2$, that we will be considering in our simulations,
 $\beta_1=0.08360$ and $\beta_2=-0.01295$. For details on how to compute the shape coefficients we refer to Ref.~\cite{Hasenfratz:1989pk}.
In Figure~\ref{fig:sigmaeff} we show the result of the ratio $\Sigma_{\rm eff}/\Sigma$ as a function of $M_{ss}^2/F^2$, for $N_s=2$, $F=90$ MeV and  $L_6^r(M_\rho)=0.07 \cdot 10^{-3}$. The NLO corrections are quite significant, up to 30$\%$ for the masses considered.

Concerning the quenched limit of the zero-mode integral over $U(N_l)$, a prescription using the supersymmetric  or replica methods gives the well-known result for the quenched partition functional~\cite{Splittorff:2002eb,Fyodorov:2002wq} that matches quenched RMT (qRMT). The  low-lying eigenvalues of the Dirac operator should then follow the predictions of qRMT. Comparing the eigenvalues computed numerically in this PQ setup with the predictions of qRMT, we can extract  $\Sigma_{\rm eff}$ of eq.~(\ref{eq:sigmaeff}). If we do this for different values of the sea-quark mass, we can study the sea-quark mass dependence of $\Sigma_{\rm eff}$, from which we can in principle disentangle $\Sigma$ and $L^r_6(\mu)$, up to NNLO corrections (assuming we have an independent  determination of $F$). 

A relevant question is however what are the eigenvalues that should be matched to RMT. Since there is a cutoff over which the ZMChPT should not be a good description, we expect that when the eigenvalues roughly reach such cutoff they should get significant corrections from the massive modes and therefore the matching to RMT should break up. A rough estimate would be the condition that 
$\lambda \leq m_{th}$, where $m_{th}$ is the value of the quark mass corresponding to the $p$-regime. For example taking the value
of $m_{th}$ such that $M L \geq 3$,  converts into the condition $\lambda \Sigma V \leq 9 F^2 L T/2$, which is roughly $6-7$ for our lattices. This results in the expectation that only the few lowest eigenvalues ($<3-4$ for our lattices) are below the threshold. For the largest eigenvalues , deviations from RMT could be sizeable, and the associated systematic uncertainty should be reduced by simulating at larger volumes.

\subsection{Random matrix theory}\label{sec:rmt}

We consider  the gaussian chiral unitary model described by the partition function
\begin{equation}
\mathcal{Z}_\nu(\hat{m}_1,...,\hat{m}_{N_f})=\int dW e^{-\frac{N}{2}{\rm Tr}(W^\dagger W) }\prod_{i=1}^{N_l} {\rm det} (\hat{D}+\hat{m}_i),
\end{equation}
where
\begin{equation}
\hat{D}= \left(
\begin{array}{cc}  0 & W \\
-W^\dagger&  0\\
\end{array}\right),
\end{equation}
and $W$ is a complex rectangular matrix of dimensions $(N+\nu)\times N$. Here $N$ plays the role of the space-time volume, multiplied by a constant. We are interested in the large-$N$ scaling limit at fixed $\mu_i=\hat{m}_iN$. The partition function $\mathcal{Z}_\nu$ then provides  an equivalent description of the zero mode-chiral theory partition function in eq.~(\ref{eq:shuryak}) \cite{Shuryak:1992pi,Verbaarschot:1993pm,Verbaarschot:1994qf,Verbaarschot:2000dy} with $N_l$ flavours of mass $\hat{m}_i$ and fixed topological charge $\nu$, with the identification $N\hat{m}_i=m_i\Sigma V=\mu_i$.

If $x$  is the $k$-th smallest eigenvalue of the matrix $\sqrt{W^\dagger W}$, the probability distribution associated to the microscopic eigenvalue $\zeta=Nx$ can be written as
\begin{equation}
p_k^\nu(\zeta;\{\mu\})= \int_0^{\zeta}d\zeta_1 \int_{\zeta_1}^{\zeta}d\zeta_2 \dots\int_{\zeta_{k-2}}^{\zeta} d\zeta_{k-1} \omega_k^\nu
(\zeta_1,\dots,\zeta_{k-1},\zeta;\{\mu\}),
\end{equation}
with $0\leq \zeta_1\leq \dots\leq \zeta_{k-1}\leq \zeta$. The explicit form of $\omega_k^\nu
(\zeta_1,\dots,\zeta_{k-1},\zeta;\{\mu\})$ is known in the microscopic limit \cite{Nishigaki:1998is,Damgaard:2000ah}.
For instance, in the quenched case $N_l=0$ one has
\begin{equation}
\omega_{kq}^\nu
(\zeta_1,\dots,\zeta_{k-1},\zeta_k;\{0\})=W_k^\nu e^{-\zeta_k^2/4}\zeta_k\prod_{i=1}^{k-1}\zeta_i^{2\nu+1}\prod_{k-1\geq i >j\geq 1}s(\zeta_i,\zeta_j)^4\times
\end{equation}
$$
Z_2(\underbrace{s(\zeta_k,\zeta_1),s(\zeta_k,\zeta_1),\dots,s(\zeta_k,\zeta_{k-1}),s(\zeta_k,\zeta_{k-1})}_{2(k-1)},\underbrace{\zeta_k,\dots,\zeta_k}_\nu),
$$
with
\begin{equation}
s(\zeta_i,\zeta_j)=\sqrt{\zeta_i^2-\zeta_j^2},
\end{equation}
and
\begin{eqnarray}
Z_2(s_1,\cdots,s_n) =  \frac{{\rm det}A^{(n)}}{\Delta^{(n)}},\nonumber
\end{eqnarray}
\begin{eqnarray}
\left[A^{(n)}\right]_{ij}  =  s_i^{j-1}I_{j+1}(s_i),\;\;
\Delta^{(n)} =  \prod_{n\geq i >j\geq 1}(s_i^2-s_j^2),
\end{eqnarray}
where $I_i$ are modified Bessel functions. The coefficient $W_k^\nu$ can be fixed such that the probability $p_k^\nu$ is normalized to unity. There is an interesting property, called flavour-topology duality, which manifests itself at zero mass: $p_k^\nu(\zeta;\{0\})$ depends on the number of dynamical flavours and the topological charge only through the combination $\xi=N_l+|\nu|$. 
The microscopic spectral density 
\begin{equation}
\rho_S^\nu(\zeta;\{\mu\})=\sum_{k=1}^\infty p_k^\nu(\zeta;\{\mu\})
\end{equation}
coincides by construction with the one computed in the ZMChT \cite{Damgaard:1998xy}. For instance, the quenched LO spectral density is given by
\begin{equation}
\rho_{Sq}^\nu(\zeta;0)=\frac{\zeta}{2}\left[J_\nu(\zeta)^2-J_{\nu+1}(\zeta)J_{\nu-1}(\zeta)            \right].
\end{equation}
The equivalence can be extended to generic $n$-point density correlation functions. 
It is possible to show that probability distributions of single eigenvalues can be defined also in the chiral effective theory by means of recursion relations involving  all spectral correlators \cite{Akemann:2003tv}. The clear advantage of RMT is that the probability distributions are computable in practice, while in the chiral effective theory explicit expressions are missing.    
By assuming this equivalence holds for all spectral correlators, it is then legitimate to match the expectation values of the low eigenvalues of the massless QCD Dirac operator $\lambda_k$ with the predictions of  the corresponding RMT. 

\vspace{0.5cm}

We will be considering here a situation where two flavours of degenerate sea quarks have masses that are sufficiently large to be in the $p$-regime. In this case the sea quark mass does not appear explicitly in the ZMChT/RMT, as we have discussed. The latter corresponds to a theory with $N_l \rightarrow 0$ light flavours, that is {\it quenched} RMT (qRMT). The sea quark mass
dependence comes in only through $\Sigma_{\rm eff}(M_{ss})$ and can be predicted at NLO, as explained in the previous section. Therefore  we expect 
\begin{equation}\label{rmt_matching}
\langle \zeta_k \rangle^\nu_{\rm qRMT}   =\left. \Sigma_{\rm eff}(M_{ss})\right|_{N_l=0} V \langle\lambda_k\rangle ^\nu_{\rm QCD}(M_{ss})     ,
\end{equation}
where $M_{ss}$ is the sea pion mass, $\left. \Sigma_{\rm eff}(M_{ss}) \right|_{N_l=0}$ is given in eq.~(\ref{eq:sigmaeff}), and expectations values are computed in RMT as
\begin{equation}
\langle \zeta_k \rangle^\nu_{\rm qRMT} = \int \dif\zeta\, p_k^\nu(\zeta;0) \zeta.
\end{equation}
In this matching we assume that the QCD quark masses, the eigenvalues of the Dirac operator and the quark condensate are properly renormalised. 
The prediction for the ratio $\langle \zeta_k \rangle^\nu_{\rm qRMT}/\langle \zeta_l \rangle^\nu_{\rm qRMT}$ is parameter-free and can be compared directly with $\langle\lambda_k\rangle ^\nu_{\rm QCD}/\langle\lambda_l\rangle ^\nu_{\rm QCD}$ at any fixed $M_{ss}$.

On the other hand, if we consider ratios at different sea quark masses of the form
$\langle\lambda_k\rangle_{\rm QCD}^\nu(M_1)/ \langle\lambda_k\rangle_{\rm QCD}^\nu(M_2)$, 
($M_{1,2}$ are two different sea pion masses), we can assume that they can be matched 
to qRMT with appropriate values $\Sigma_{\rm eff}(M_{1,2})$ of the effective chiral
condensate. Therefore
\begin{gather}
\label{eq:ev_condensate}
\frac{\langle\lambda_k\rangle_{\rm QCD}^\nu(M_1)}{\langle\lambda_k\rangle_{\rm QCD}^\nu(M_2)}
= \frac{\langle\zeta_k\rangle^\nu_{{\rm qRMT}}}{\langle\zeta_k\rangle^\nu_{{\rm qRMT}}}\,
\left. \frac{\Sigma_{\rm eff}(M_2)}{\Sigma_{\rm eff}(M_1)} \right|_{N_l=0}
=\left. \frac{\Sigma_{\rm eff}(M_2)}{\Sigma_{\rm eff}(M_1)}\right|_{N_l=0}\,.
\end{gather}
It follows that information on the mass dependence of $\Sigma_{\rm eff}$,
and hence on $L_6$, can be obtained from suitable eigenvalue ratios.

%% file: num.tex
\section{Results on Dirac spectral observables}
\label{sec:num}

\begin{table}
\begin{center}
\begin{tabular}{|c|c|c|ll|}
\hline
\multicolumn{5}{|c|}{} \\[-0.3cm] 
\multicolumn{5}{|c|}{$\beta= 5.3$, $c_{\rm sw}=1.90952$, $V/a^4=48\times 24^3$} \\ [0.1cm]
\hline
& & & & \\[-0.3cm]
label & $\kappa$ & $aM_{ss}$ &  \multicolumn{2}{|c|}{$N_{\rm cfg}$} \\[0.1cm]
\hline
& & & & \\[-0.3cm]
D$_4$ &  0.13620 & 0.1695(14)  &  156  &  \\ 	  
D$_5$ & 0.13625  &  0.1499(15)  & 169 &\\	 
D$_6$ & 0.13635 & 0.1183(37)   &	 246 &(D$_{6\rm{a}}$: 159, D$_{6\rm{b}}$: 87)  \\[0.1cm]
\hline
\end{tabular}
\end{center}
\caption{Simulation parameters.} \label{table:sim}
\end{table}
We have carried out our computations on CLS lattices of size $48\times 24^3$.
The configurations have been generated with non-perturbatively $\Oa$ improved Wilson
fermions at $\beta=5.3$ and sea quark masses given by $\kappa=0.13620, 0.13625, 0.13635$ \cite{DelDebbio:2006cn}.
The simulations have been performed with the DD-HMC algorithm \cite{Luscher:2005rx};
further details can be obtained in~\cite{DelDebbio:2007pz}.
The lattice spacing has been determined to be
$a=0.0784(10)~$fm in \cite{DelDebbio:2006cn}, which implies that our lattices
have physical size $L\simeq 1.88$ fm
and sea pion masses of 426, 377 and 297 MeV, respectively.
However, preliminary results
from more precise determinations through different methods yield
$a\simeq 0.070~Ê$fm \cite{Brandt:2010ed}. We will consider both values in our analysis.

Following \cite{DelDebbio:2006cn}, we will refer to our three lattices as D$_4$, D$_5$ and D$_6$. It has to
be noted that for the D$_6$ lattice we have two statistically independent ensembles,
that we dub D$_{6{\rm a}}$ and D$_{6{\rm b}}$. We have analyzed 246 D$_6$
configurations, 169 D$_5$ configurations and 156 D$_4$ configurations; in all cases successive saved
configurations are separated by 30 HMC trajectories of length $\tau=0.5$. In Table \ref{table:sim} we collect the simulation parameters. The sea pion masses in lattice units are taken from \cite{DelDebbio:2007pz} for the lattices D$_4$, D$_5$, while for D$_6$ we performed an independent evaluation from the pseudoscalar correlator computed on 96 CLS configurations. The resulting value implies $M_\pi L=2.84(9)$, which complies with the
stability bound for the simulation algorithm derived in~\cite{DelDebbio:2005qa}.
On the other hand, $M_\pi L < 3$ implies sizeable finite volume effects
in p-regime physics, which in the present work are accounted for within ChPT.

\begin{figure}[t!]
\begin{center}
\hspace*{-1mm}\includegraphics[width=7.1cm]{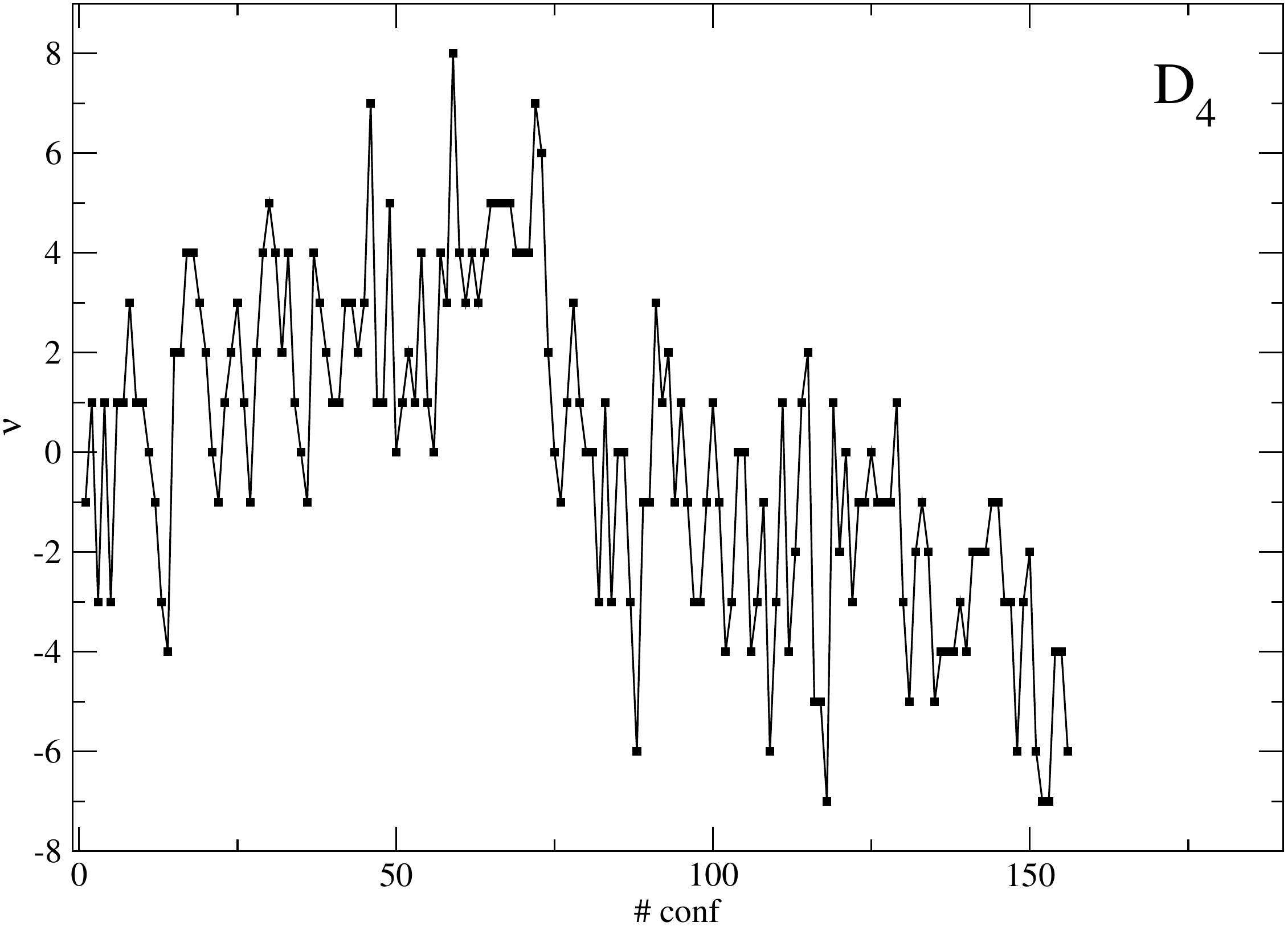}
\hspace{7mm}\includegraphics[width=7.1cm]{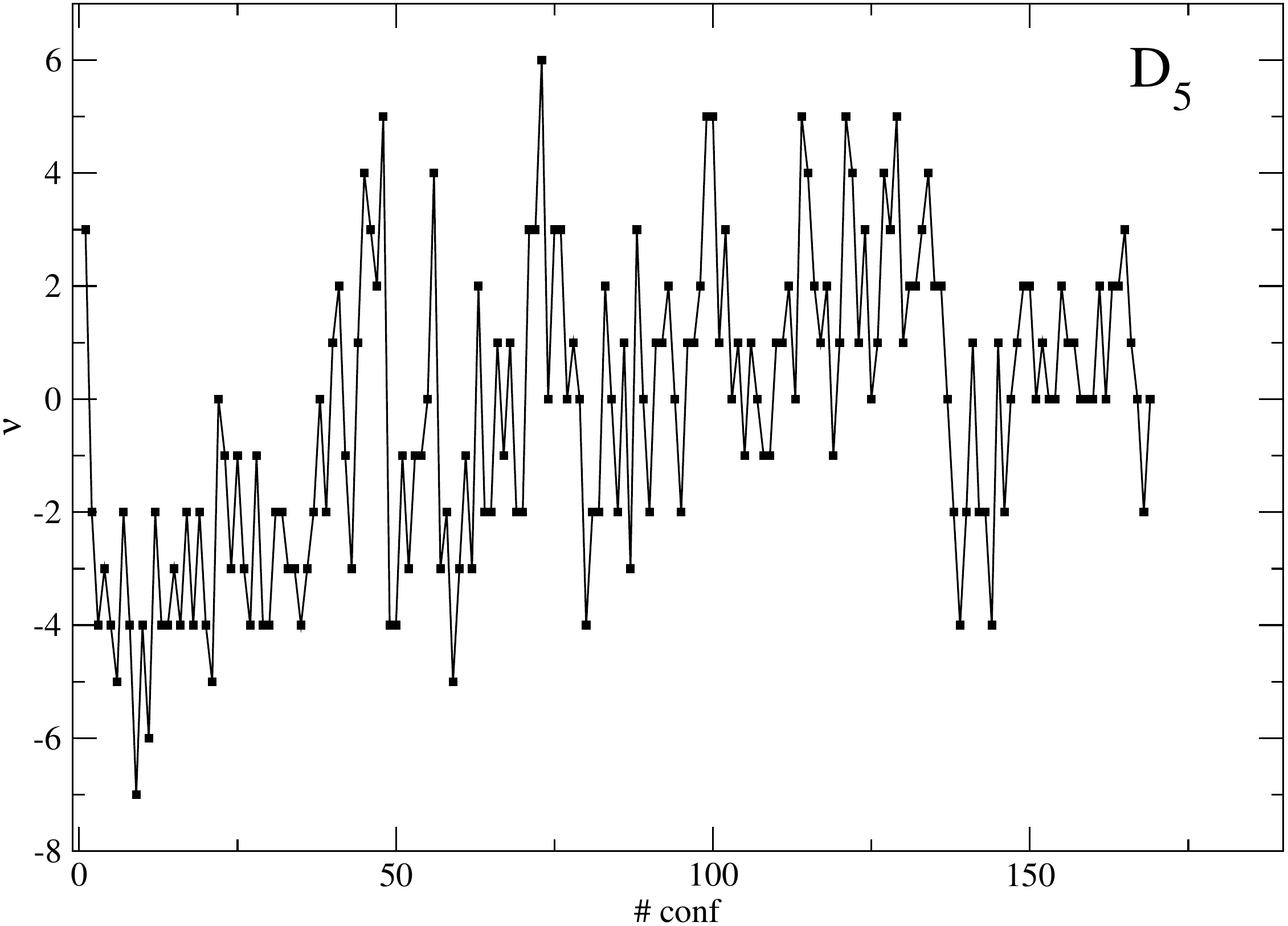}

\vspace*{2mm}

\hspace*{1mm}\includegraphics[width=7.1cm]{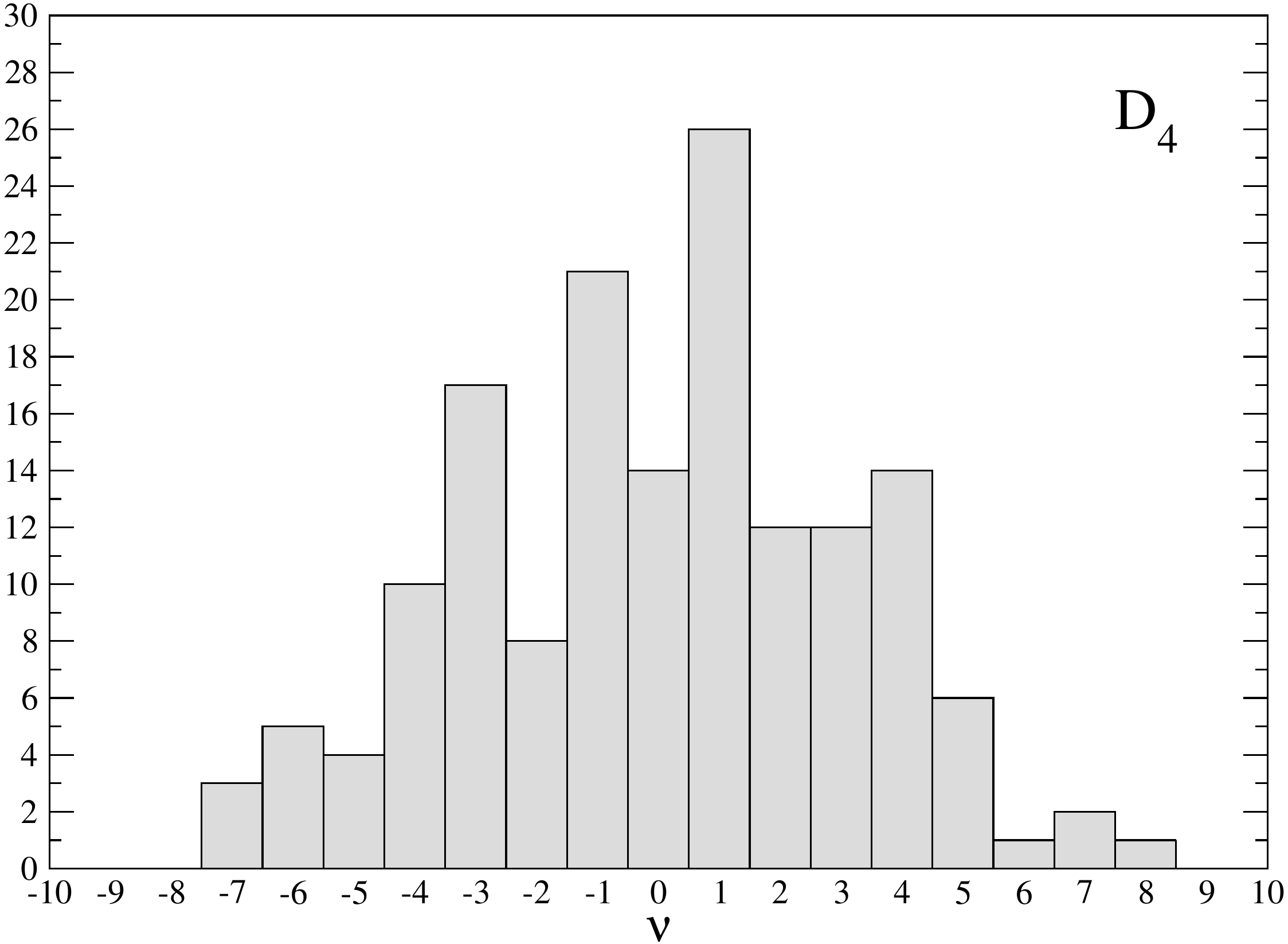}
\hspace{7mm}\includegraphics[width=7.1cm]{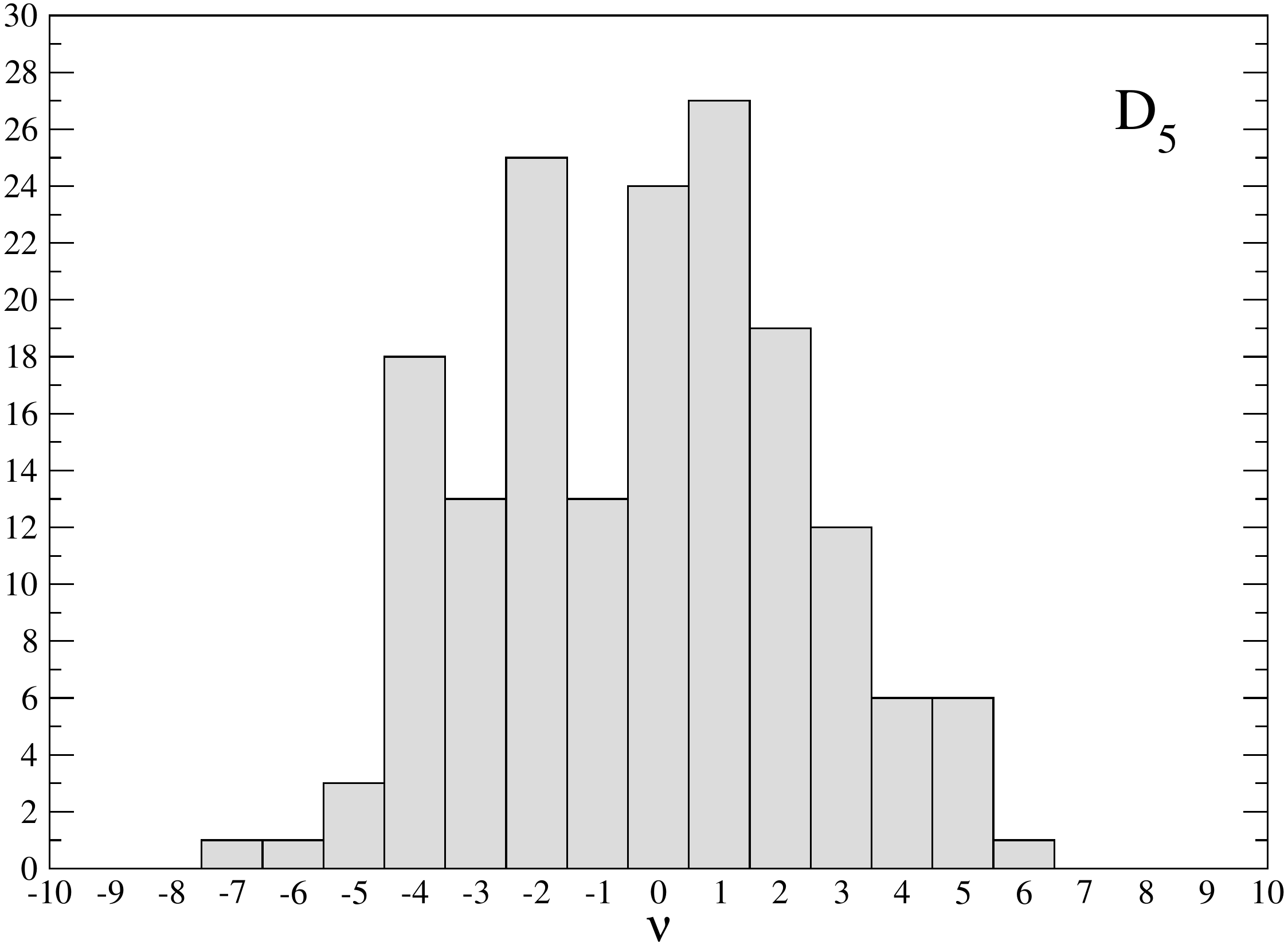}
\end{center}\vspace{-7mm}
\caption{MC history and distribution of the topological charge in lattices D$_4$ and D$_5$.}
\label{fig:topologyD4}
\end{figure}

On these configurations we have built the massless Neuberger-Dirac operator \cite{Neuberger:1997fp,Neuberger:1998wv}
\begin{equation}
D_{\rm N} = \frac{1}{\overline{a}}\left\{1+\gamma_5 \;{\rm sign} (Q)   \right\},
\label{D_Ndef}
\end{equation}
with
\begin{equation}
Q=\gamma_5(aD_{\rm W}-1-s),\;\;\;\;\overline{a}=\frac{a}{1+s},
\label{Qdef}
\end{equation}
where $D_{\rm W}$ is the Wilson Dirac operator. The parameter $s$ governs the locality of $D_N$ and has been fixed to $s=0.4$ for all our simulations.
A discussion on the locality properties of the Neuberger-Dirac operator in our setup can be found in Appendix~\ref{app:locality}.

Our Neuberger fermion code is the same used in previous quenched studies~\cite{Giusti:2003gf,Giusti:2004yp,Giusti:2006mh,Giusti:2007cn,Giusti:2008fz},
and is designed specifically to perform efficiently in the
$\epsilon$-regime~\cite{Giusti:2002sm}.  
Our data analysis methods, including a discussion of autocorrelations in the observables under consideration, are discussed
in Appendix~\ref{app:analysis}.

\begin{figure}[t!]
\begin{center}
\includegraphics[width=7.1cm]{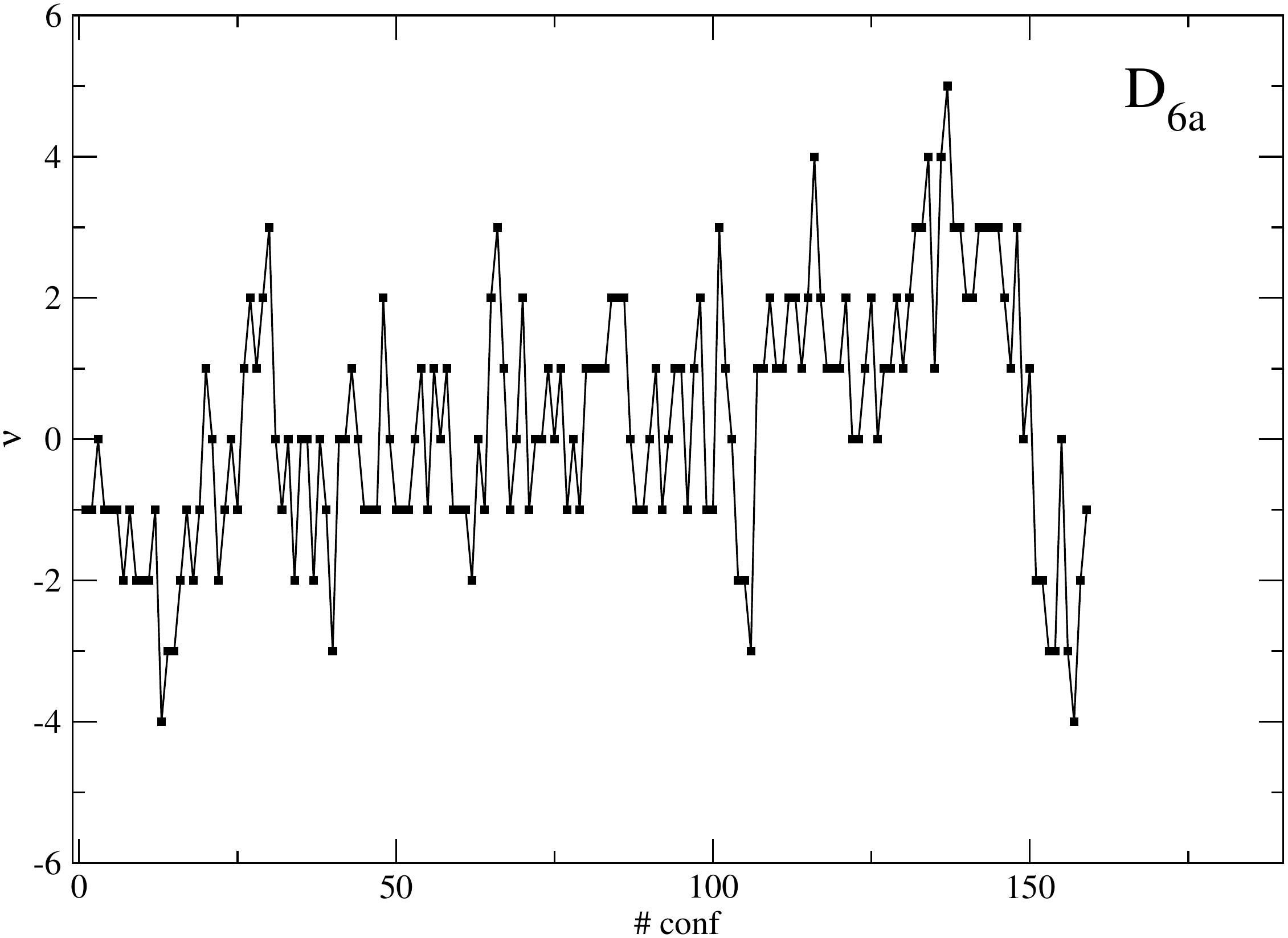}
\includegraphics[width=7.2cm]{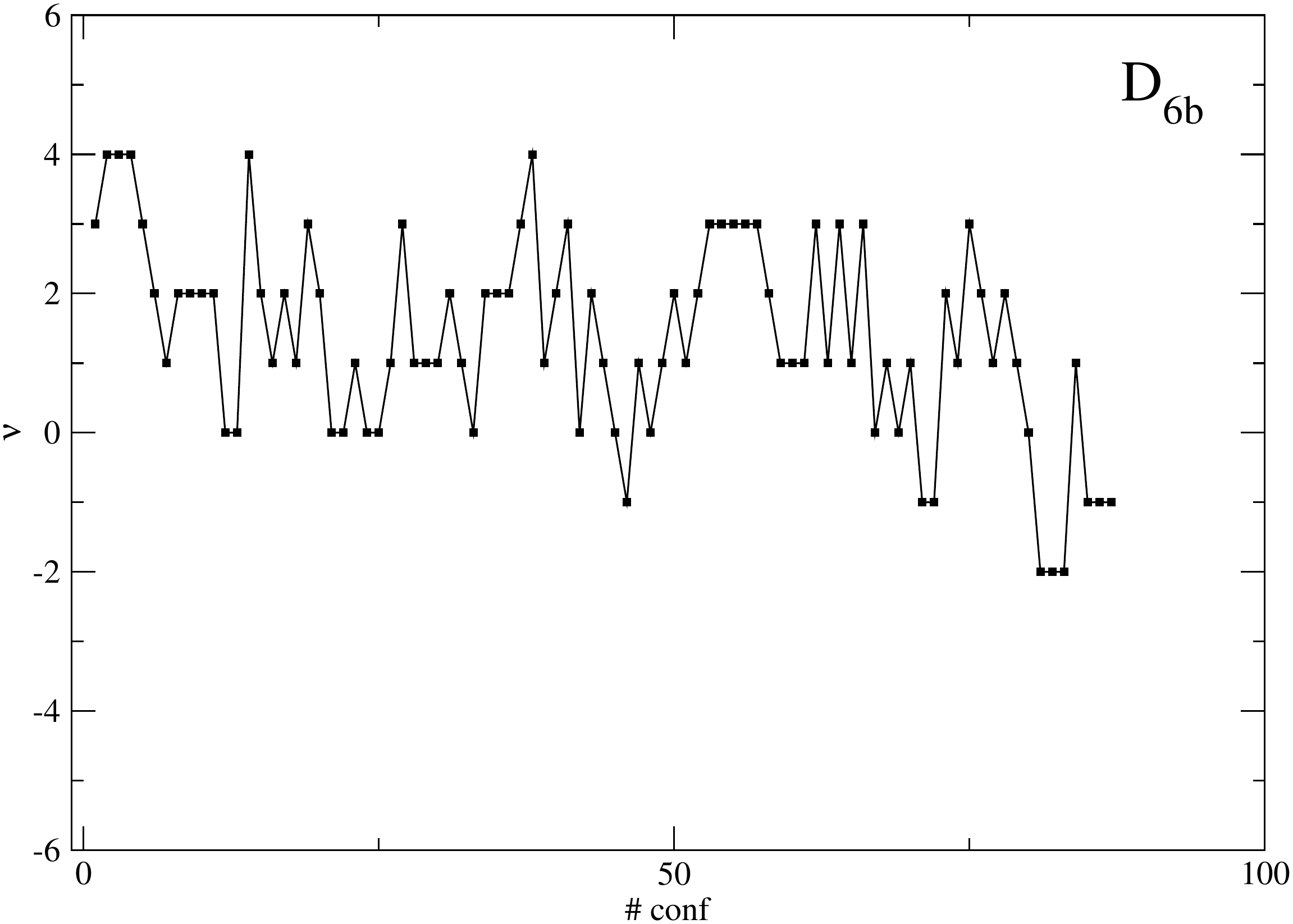}

\vspace*{2mm}

\includegraphics[width=7.1cm]{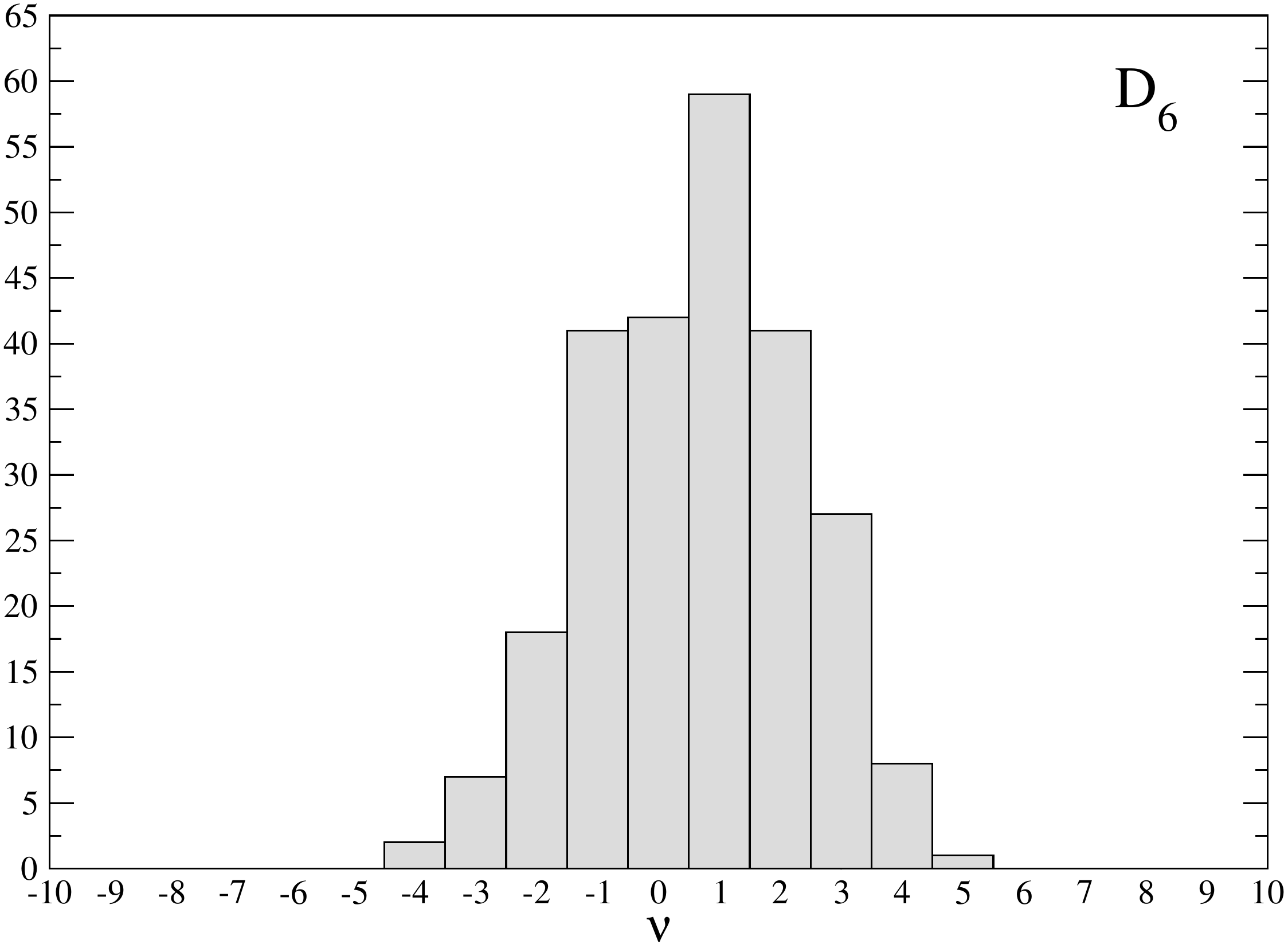}
\end{center}\vspace{-7mm}
\caption{MC history and distribution of the topological charge in lattice D$_6$.}
\label{fig:topologyD6}
\end{figure}

\begin{table}[h!]
\begin{center}
\begin{tabular}{|c|c|c|}
\hline
lattice &  $\langle\nu\rangle$ & $\langle\nu^2\rangle$ \\
\hline
D$_4$ & 0.01(55) & 9.9(1.5) \\
D$_5$ & -0.24(40) & 6.93(98) \\
D$_6$ & 0.62(24) & 3.36(47) \\
\hline
\end{tabular}
\end{center}
\caption{Results for the expectation value of the topological charge and its square.} 
\label{table:nu}
\end{table}

\subsection{Topological charge}

A first, immediate application of having constructed the Neuberger-Dirac operator
$D_{\rm N}$ on a given dynamical configuration is the determination
of the topological charge of the latter by computing the index of $D_{\rm N}$,
\begin{equation}
\nu=n_+ -n_-
\end{equation}
where $n_+$ ($n_-$) is the number of zero modes of $D_{\rm N}$ with positive (negative) chirality.

In the upper part of Figs. \ref{fig:topologyD4}--\ref{fig:topologyD6}
we show the Monte Carlo history of the topological charge for our three lattices. The topology sampling proceeds smoothly, although
there are clear hints at the presence of sizeable autocorrelations (cf. Appendix~\ref{app:analysis}). The histograms in the lower panels show the distribution of the
measured topological charges, which qualitatively exhibits the expected Gaussian-like shape and width.
This finding is consistent with the study reported in~\cite{Schaefer:2010hu}, since our computations
take place at a value of the lattice spacing sufficiently larger than the threshold
$a \lesssim 0.05~\fm$ below which topology is expected to exhibit freezing symptoms.

In Table~\ref{table:nu} we quote our results for the expectation values
$\langle\nu\rangle$ and $\langle\nu^2\rangle$.

\subsection{Low modes of the Dirac operator}
We have computed the 10 lowest eigenvalues of the Neuberger-Dirac operator on lattices D$_4$, D$_5$ and D$_6$ by adopting the numerical techniques described in~\cite{Giusti:2002sm}.

The eigenvalues of $D_{\rm N}$ appear in general in complex conjugated pairs and lie on a circle in the complex plane
\begin{equation}
D_{\rm N} \psi= \gamma \psi,\;\;\;\;\gamma=\frac{1}{\overline{a}}(1-e^{i\phi}).
\end{equation}
In order to compare them with the predictions of RMT, we have computed the projection \cite{Giusti:2003gf}
\begin{equation}
\lambda=\sqrt{\gamma\gamma^*}=\frac{1}{\overline{a}}\sqrt{2(1-\cos\phi)}.
\end{equation}
We have evaluated expectation values at fixed absolute value of the topological charge $|\nu|$. In Fig. \ref{fig:lambda} we show the bare eigenvalues for $|\nu|=0,1,2$.

Since the matching with RMT involves the parameter $\Sigma$, it is useful to first consider ratios of eigenvalues. 
In our case, following eq. (\ref{rmt_matching}), the QCD ratios $\langle\lambda_k\rangle^{\nu_1}/\langle\lambda_l\rangle^{\nu_2}$ can be directly matched with the qRMT predictions $\langle\zeta_k\rangle^{\nu_1}/\langle\zeta_l\rangle^{\nu_2}$. 
In Tables \ref{tab_res:ratios1}, \ref{tab_res:ratios2}~(App.~\ref{app:analysis}) we report the results for eigenvalue ratios involving the four lowest-lying eigenvalues and topologies $|\nu|=0,1,2$, together with qRMT predictions. 
It should be pointed out that  the matching to RMT should work provided $\lambda_k \Sigma_{eff} V$ is not much larger that 1. For the lattice parameters we are considering, we set out cutoff at $k \leq 4$ for which the parameter is below 10.  Since $k=4$ is probably borderline, we will not include it in the extraction of $\Sigma_{eff}$ though.

\begin{figure}
\begin{center}
\includegraphics[width=13cm]{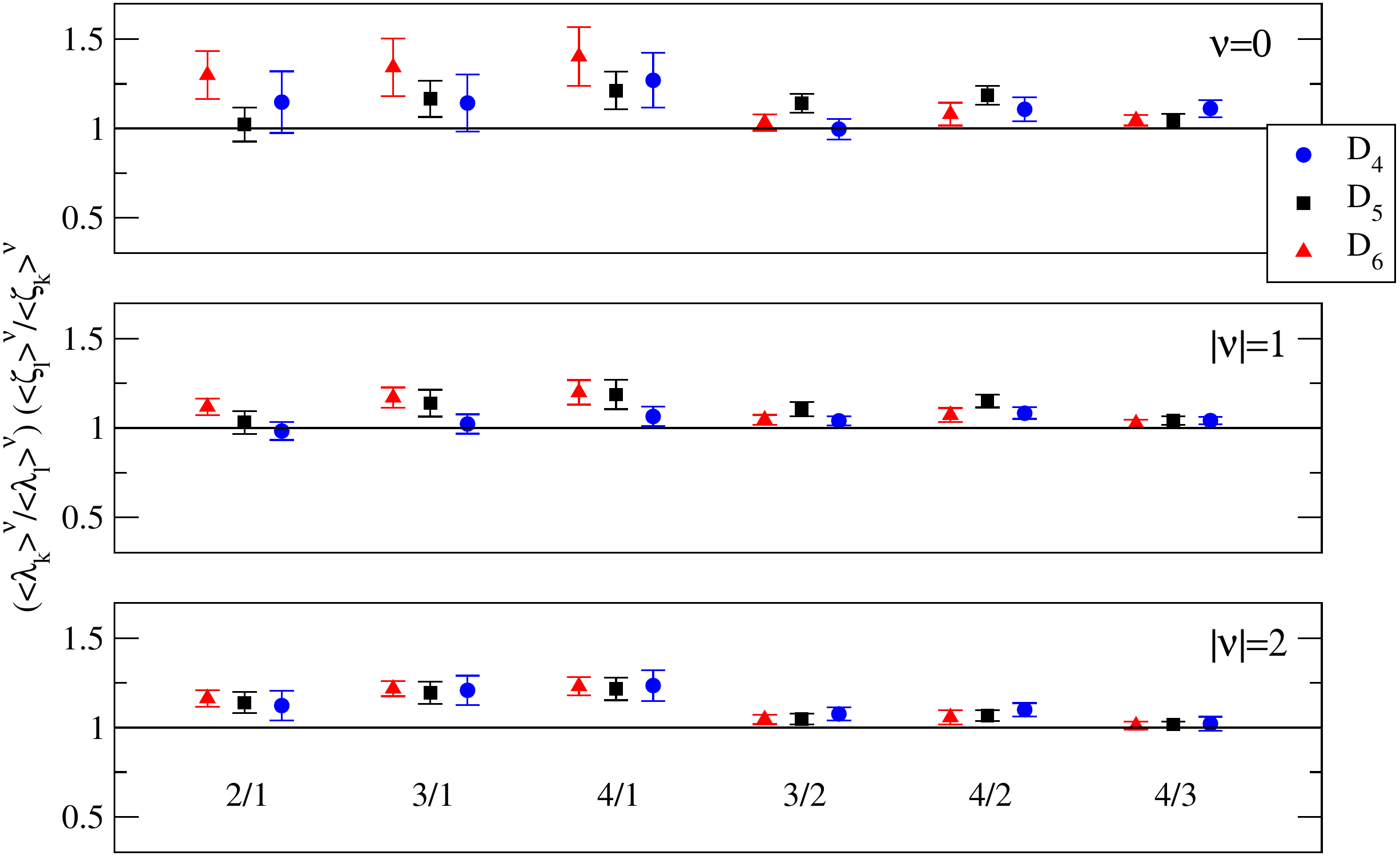}
\caption{Eigenvalue ratios at fixed $|\nu|$, normalized to qRMT predictions, for the indices $(k,l)=(2,1), (3,1), (4,1), (3,2), (4,2), (4,3)$.}\label{fig:ratios}
\end{center}
\end{figure}

The ratios at fixed topological charge are shown in Figs. \ref{fig:rat_nu0}, \ref{fig:rat_nu1}, \ref{fig:rat_nu2}~(App.~\ref{app:analysis}); moreover, in Fig. \ref{fig:ratios} we report the ratios 
$\langle\lambda_k\rangle^{\nu}/\langle\lambda_l\rangle^{\nu}$ normalized to the corresponding qRMT predictions, for $|\nu|=0,1,2$ and for several combinations $k,l$ given in the bottom of the plot.
This allows to appreciate clearly the precision and level of agreement with qRMT of each
specific case.
Finally, ratios at fixed $k$ involving different topological sectors are presented in Fig. \ref{fig:rat_k}~(App.~\ref{app:analysis}).

While the RMT prediction seems to work well for
ratios not involving $\lambda_1$, the ratios $\langle\lambda_k\rangle/\langle\lambda_1\rangle$ exhibit 
somewhat more significant deviations.  On the other hand, ratios
between eigenvalues in different topological sectors follow well RMT predictions
also in the case of $\lambda_1$, as shown in \refig{fig:rat_k}, albeit with larger
errors. The data presented in this work
do not allow for a full assessment of the systematics of these deviations, as this would require e.g. further values of the lattice spacing and/or physical volume.
It is worth noting however that there are no clear differences between the three lattices,
which we can take as an indication that corrections associated with relatively small
values of $m_s$ are small. Concerning finite lattice spacing effects, having an
estimate of the associated corrections in Wilson ChPT~\cite{Sharpe:1998xm,Rupak:2002sm,Bar:2003mh} would be welcome,
although our value of the lattice spacing is quite small. Finally, one has to keep
in mind that the impact of autocorrelations on statistical errors
cannot be estimated accurately for our ensembles.
While we have attempted to stay on the safe side by quoting
conservative errors that ought to include autocorrelations properly, it cannot be
excluded that some errors are underestimated.
Details are provided in App.~\ref{app:analysis}.

\begin{figure}[!t]
\begin{center}
\includegraphics[width=10cm]{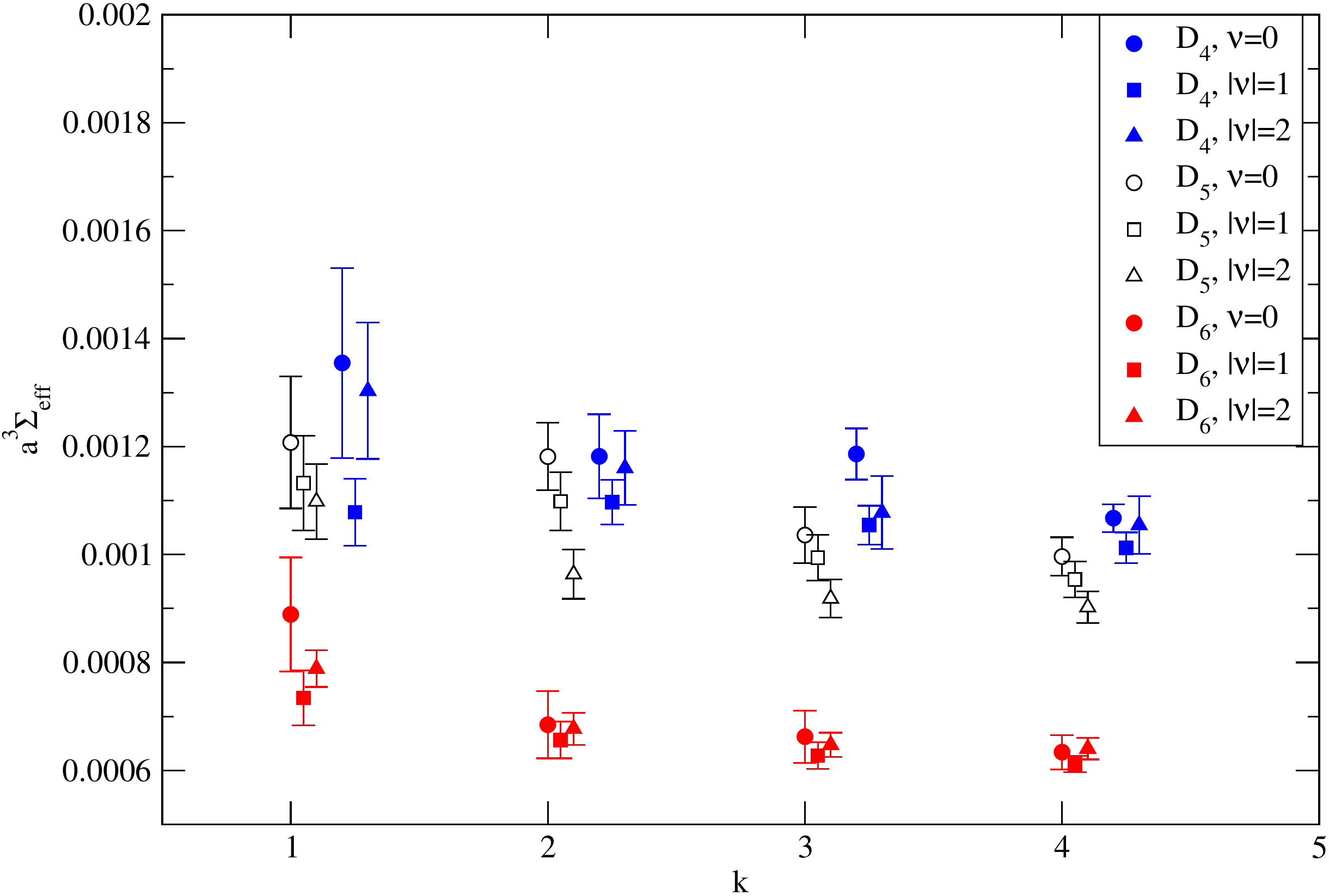}
\caption{The bare effective condensate $\Sigma_{\rm eff}$ extracted from the matching in eq.~(\protect\ref{sigmaeff_mat}), for $k=1,2,3,4$ and $|\nu|=0,1,2$. The data for D$_4$ have been shifted in the horizontal axis for better clarity.}\label{fig:sigmaeffres}
\end{center}
\end{figure}

\subsection{Effective quark condensate}

In the spirit of the mixed regime ChiPT analysis, our data also allow to study
the mass dependence of the effective condensate, cf. eq.~(\ref{eq:ev_condensate}).
In Table \ref{tab:sigmaeff} we report the values of the bare effective condensate extracted from the matching
\begin{equation}\label{sigmaeff_mat}
\Sigma_{\rm eff}(M_{ss})=\frac{\langle \zeta_k \rangle^\nu_{\rm qRMT}}{V \langle\lambda_k\rangle ^\nu_{\rm QCD}}
\end{equation}
for $k=1,2,3,4$ and $|\nu|=0,1,2$. The results are shown in Fig. \ref{fig:sigmaeffres}, where one can 
observe that, at fixed value of the sea quark mass, $\Sigma_{\rm eff}$ does not depend on $k$ and $\nu$ within the statistical precision (with larger errors for $k=1$). By averaging over  $k=2,3$ and $|\nu|=0,1,2$ we obtain
\begin{eqnarray}
a^3\Sigma_{\rm eff}&=& 0.00113(3)(4) \;\;\; ({\rm D}_4),\nonumber\\
a^3\Sigma_{\rm eff}&=& 0.00103(3)(4) \;\;\; ({\rm D}_5), \\
a^3\Sigma_{\rm eff}&=& 0.00066(2)(5) \;\;\; ({\rm D}_6).\nonumber
\end{eqnarray}
The first error is the statistical one, while the second uncertainty is a systematic effects estimated by adding the values for $k=1$ in the average. We have checked that including $k=4$ in the fit does not change the values within the statistical accuracy but decreases slightly the errors.

Finally, in Fig. \ref{fig:sigmarat}~(App.~\ref{app:analysis}) we show the ratios defined in eq.~(\ref{eq:ev_condensate}) for $k=1,2,3,4$ and $|\nu|=0,1,2$: they can be matched to the ratios of $\Sigma_{\rm eff}$ at different quark masses. Here we can see that, as expected, those ratios do not depend on the topology and on $k$, within the statistical uncertainties.

%% file: fits.tex
\section{Fits to NLO Chiral Perturbation Theory}
\label{sec:fits}

On the basis of the evidence presented in the previous section, now we assume that the matching to ChPT in the mixed-regime works in this range of sea quark masses and volumes, and  try to extract the low-energy couplings from the sea-quark mass dependence of the two quantities $\Sigma_{\rm eff}$ and $\langle\nu^2\rangle$.

The NLO predictions from ChPT are summarized in eqs.~(\ref{eq:sigmaeff}) and (\ref{eq:nu2nlo}). As expected they depend on the two leading order LECs, $\Sigma$ and $F$, but also on the ${\mathcal O}(p^4)$ couplings $L_6$, $L_7$ and $L_8$. 

\begin{figure}
\begin{center}
\includegraphics[width=9cm]{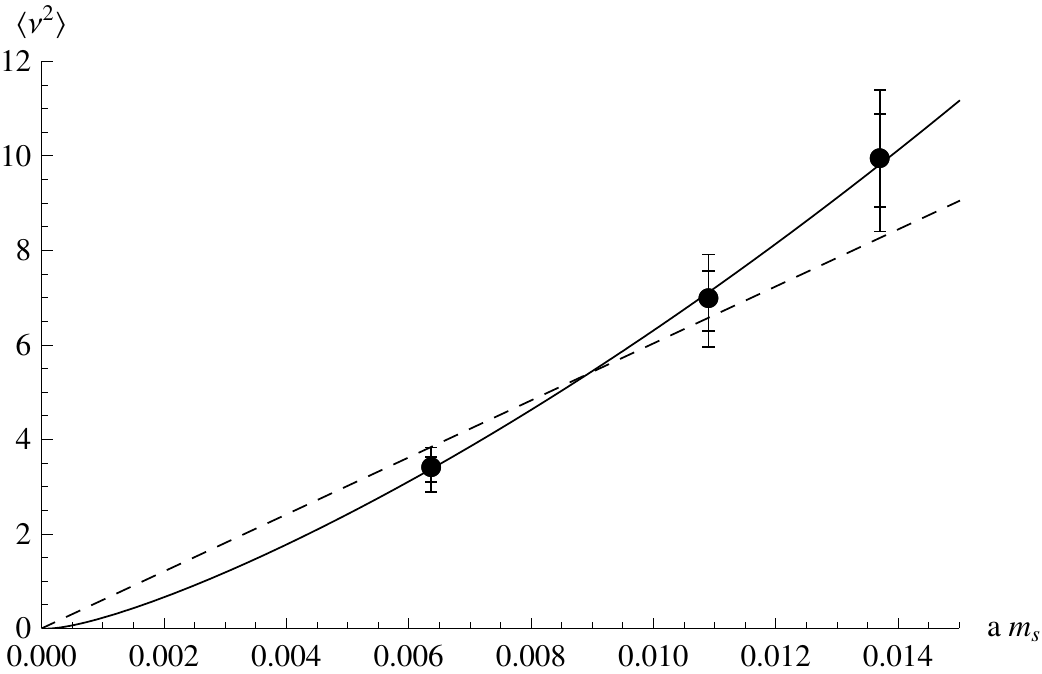}
\caption{$\langle \nu^2\rangle$ versus the sea quark mass $a m_s^{\overline{\rm MS}}(2{\rm GeV})$. The smaller errors are the statistical ones and the largest include our estimate of autocorrelations. The dashed and solid lines correspond to the best LO and NLO ChPT fits respectively (for $a=0.078$~fm).}
\label{fig:nu2}
\end{center}
\end{figure}

\begin{table}
\begin{center}
\begin{tabular}{|c|c|c|}
\hline
& & \\ [-0.4cm]
 Lattice  & $a m$ & $a m_R({\overline{\rm MS}},2\;{\rm GeV})$ \\
\hline
D$_4$ &   0.00954(8) & 0.01366(23) \\
D$_5$ &   0.00761(7) & 0.01090(19) \\
D$_6$ &   0.00445(22)& 0.00637(33) \\
\hline
\end{tabular}
\end{center}
\caption{Bare and renormalised  ($\overline{\rm MS}$ scheme at 2 GeV) PCAC sea quark masses.}
\label{tab:pcac}
\end{table}
We first consider the topological charge distribution. The statistical error in this quantity is fairly large, but it is encouraging to see that there is a very clear dependence on the sea quark mass as shown in Fig.~\ref{fig:nu2}. We have fitted both to the full NLO formula in
eq.~(\ref{eq:nu2nlo}), and to the linear LO behaviour. In either case $m_s$ is taken to be the PCAC Wilson mass renormalised in the
$\overline{\rm MS}$ scheme at 2~GeV, tabulated in Table~\ref{tab:pcac}. The results for the lattice D$_4$ and D$_5$ are taken from \cite{DelDebbio:2007pz}, while we have computed that of D$_6$,\footnote{We thank A.~J\"uttner for providing the necessary Wilson propagators for D$_6$.} using the renormalization constants and improvement coefficients  of \cite{DellaMorte:2005se,DellaMorte:2008xb,
DellaMorte:2005kg,DellaMorte:2008ad,Fritzsch:2010aw}.

At LO the slope provides a direct measurement of $\Sigma$ in the same scheme.
At NLO we fit for $\Sigma$ and the combination $\left[L^r_8 + 2 (L^r_6+L^r_7)\right](M_\rho)$ after fixing $\mu=M_\rho$ and rewriting
$M_{ss}=2m_s\Sigma/F^2$. The value of $F$ is fixed to 90~MeV; the systematic uncertainty related to this choice is estimated by varying
$F$ by $\pm 10$~MeV.

In Table~\ref{tab:nu2fits} we show the results of the LO and NLO fits. In the case of the NLO,  there is a slight difference when the scale
is taken to be $a=0.078~Ê$fm \cite{DelDebbio:2006cn} or the preliminary result $a\simeq 0.070~Ê$fm \cite{Brandt:2010ed}. We quote both.  The $\chi^2$ is better for the NLO fit, but  it is not possible to exclude the LO behaviour without decreasing our statistical errors.  In physical units we get for $a=0.078-0.070~$fm:
\begin{eqnarray}
\Sigma^{\overline{\rm MS}}(2\; {\rm GeV})  = \left[262^{(33)(4)}_{(34)(5)} ~Ê{\rm MeV}\right]^3 \;- \; \left[287^{(35)(5)}_{(36)(7)} ~Ê{\rm MeV}\right]^3,\;\;\;  
\end{eqnarray}
where the first error is coming out of the fit and the second is the effect of changing $F$. 
\begin{table}
\begin{center}
\begin{tabular}{|c|c|c|c|}
\hline
& & &\\[-0.4cm]
    & $\Sigma^{\overline{\rm MS}}(2\;{\rm GeV}) a^3$ &  $[L^r_8+2 (L^r_6+L^r_7)](M_\rho)$ & $\chi^2/dof$\\
\hline
LO &   0.00182(16) & - & 1.2\\
NLO (a=0.078~Êfm) & 0.00112(48) & 0.0023(43) & 0.02 \\
NLO (a=0.070~Êfm) & 0.00106(44) & 0.0018(30) & 0.03 \\
\hline
\end{tabular}
\end{center}
\caption{Results from the chiral fits to eq.~(\protect\ref{eq:nu2nlo}).}
\label{tab:nu2fits}
\end{table}
There have been previous studies of the dependence on the topological susceptibility on the sea quark mass \cite{DeGrand:2007tx,Aoki:2007pw,Chiu:2008jq,Bazavov:2010xr}, but as far as we are aware the fits in these works have not included the NLO chiral corrections.

Let us now turn to $\Sigma_{\rm eff}$. In this case, the dependence on $m_s$ is expected starting at NLO in ChPT. The results in the previous section indicate that indeed there is a significant sea quark mass dependence in this quantity.  We perform a two-parameter NLO fit, where we fix $F$ and fit for $\Sigma$ and $L^r_6(M_\rho)$. As a function of the $\overline{\rm MS}(2\;{\rm GeV})$ sea quark mass $m_s$, we expect therefore:
\begin{eqnarray}
Z^{\overline{\rm MS}}_S \Sigma_{\rm eff}(m_s) &=& \Sigma\, \Bigg\{ 1+ { 2 m_s \Sigma  \over F^4} \Bigg[ {\beta_2 \over 2} + {1 \over 16\pi^2} \log(M_\rho V^{1/4}) + 32 L^r_6(M_\rho)
-{1 \over 16 \pi^2} \log\left({m_s \Sigma \over F^2 M_\rho^2}\right)\Bigg]
\nonumber\\
&&~~~~~~
- \frac{\beta_1}{2F^2V^{1/2}} -{2 \over F^2} g_1\left(\sqrt{{\Sigma m_s\over F^2}}, L, T\right)\Bigg\},
\end{eqnarray}
where we need the scalar density renormalization factor in the $\overline{\rm MS}$ scheme for the valence overlap fermions.
We have obtained a rough estimate of this factor by matching our valence
and sea sectors at a reference value of the pion mass, computed with mass-degenerate
quarks, following the method of~\cite{Hernandez:2001yn}. We have done this at the
unitary point on lattice D$_5$; choosing the latter instead of our lightest point
D$_6$ allows to avoid sizeable finite volume effects in the determination of pion masses.

The sea pion mass in lattice units 
is $aM_{ss}=0.1499(15)$~\cite{DelDebbio:2007pz}, while for the mass of the valence
pion at bare valence quark mass $am=0.020$ we obtain $aM_{vv}=0.153(5)$ with 63
D$_5$ configurations.\footnote{The relatively small error for this limited statistics
is a result of the use of low-mode averaging~\cite{Giusti:2004yp} in the computation of the two-point 
function of the left-handed current, from which the mass is extracted.} 
In order to obtain the renormalisation factor we then apply the matching condition
\begin{eqnarray}
\left. \left(Z^{\overline{\rm MS}}_{\rm S}\right)^{-1}  m \right|^{\rm overlap}_{M^{\rm ref}_\pi} =   \left. m^{\overline{\rm MS}}(2~\GeV)\right|^{\rm Wilson}_{ M^{\rm ref}_\pi}
\end{eqnarray}
at $aM_{\rm ref}=aM_{ss}$. 
The renormalised $\overline{\rm MS}$ mass of the sea
quark mass for the D$_5$ lattice can be read from Table~\ref{tab:pcac},
and we obtain
\begin{eqnarray}
Z^{\overline{\rm MS}}_{\rm S}(2~\GeV) = 1.84(10),
\end{eqnarray}
where the error is dominated by the one in $aM_{vv}$, i.e. in the determination of
the unitary point, and can be much improved with a larger statistics in the valence sector.
Obviously, several checks need to be done to ensure that this result is robust, such as checking the dependence on the reference pion mass, as well as on the sea quark mass. A careful study of renormalization is beyond the scope of this exploratory study, and will be performed in a forthcoming publication.

\begin{figure}
\begin{center}
\includegraphics[width=9cm]{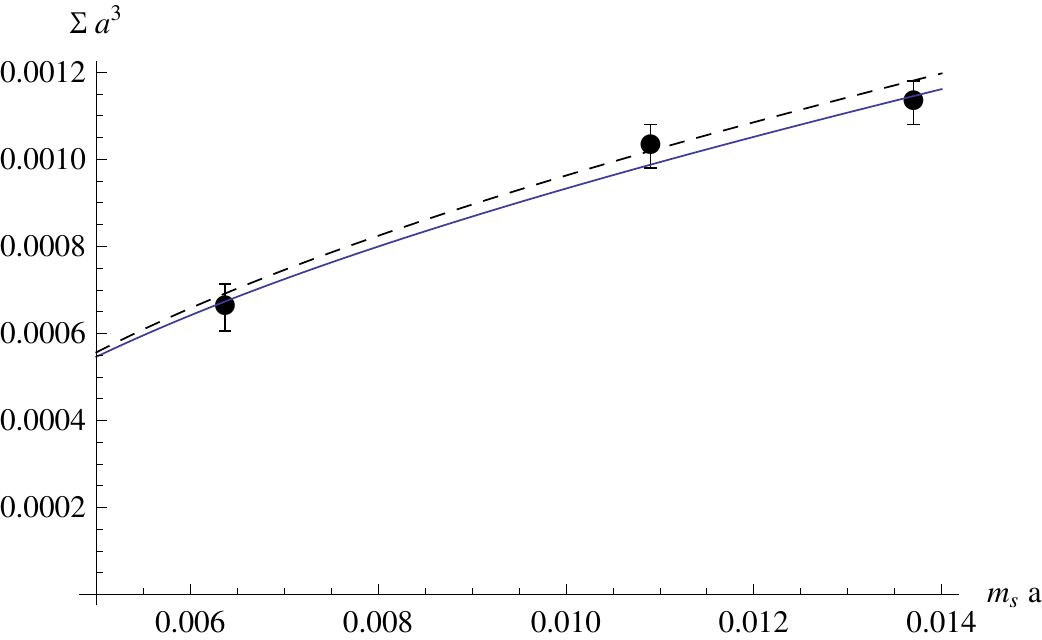}
\caption{$\Sigma_{\rm eff} a^3$ as a function of the   sea quark mass $a m_s^{\overline{\rm MS}}(2{\rm GeV})$. The dashed and solid lines correspond to the best fit for $\Sigma$ and $L_6^r(M_\rho)$ at $F=90~$MeV,  taking the scale to be $0.070~$fm and $0.078~$fm respectively.}
\label{fig:fitsigma}
\end{center}
\end{figure}

With this estimate for the renormalisation factor and fixing $F=90$ MeV, the result we obtain from the fit is, for $a=0.078-0.070~$fm 
\begin{eqnarray}
\Sigma^{\overline{\rm MS}}(2\;{\rm GeV}) a^3 &=& 0.00102(18)- 0.00099(16)\;\;\;   \\
 L_6^r(M_\rho) &=& 0.0015(11)- 0.0010(7).
\end{eqnarray}
The quality of the fit is shown in Fig.~\ref{fig:fitsigma}. Although the fit is good, it would be desirable to have more sea quark masses and smaller statistical errors to assess the systematics of this chiral fit. Particularly useful would be to test the finite-size scaling. 
Translating to physical units we have
\begin{eqnarray}
\Sigma^{\overline{\rm MS}}(2\; {\rm GeV})  = \left[255^{(14)(1)}_{(16)(4)} ~Ê{\rm MeV}\right]^3 \;- \; \left[280^{(14)(4)}_{(16)(5)} ~Ê{\rm MeV}\right]^3,\;\;\;  
\end{eqnarray}
where the only systematic error that has been estimated is that associated to the change of $F$ by $\pm 10$~MeV. 

This value of $\Sigma$  is consistent with the one obtained from the topological susceptibility above, and  both are in nice agreement with  the alternative determination of \cite{Giusti:2008vb}, that extracted the condensate from the spectrum of the Wilson-Dirac operator on  $N_f=2$ CLS configurations at the same lattice spacing and sea quark masses, but in a larger physical volume. 
A number of recent determinations of $\Sigma$  for $N_f=2$ can be found in the literature (for a recent review  see \cite{Necco:2009cq}). The matching to ChPT has been done in the $p$-regime \cite{Noaki:2008iy,Frezzotti:2008dr,Baron:2009wt}, and  also  in the $\epsilon$-regime in \cite{Lang:2006ab,DeGrand:2007tm,Fukaya:2007yv,Fukaya:2007cw,Hasenfratz:2007yj,Hasenfratz:2008ce}. Our determination uses instead a PQ mixed regime and has been obtained with significantly finer lattices than the latter. Although our result lies in the same ballpark as many of these previous determinations, it is necessary to quantify the systematic uncertainties involved in our calculation. The dispersion of results for $\Sigma$ existing in the literature is rather disturbing, and it is very important to do a proper job at evaluating the systematic uncertainties: finite $a$, systematics of the chiral fits and finite-size scaling.  

%% file: concl.tex
\section{Conclusions}

We have implemented a mixed action approach to lattice QCD in which sea quarks
are non-perturbatively $\Oa$ improved Wilson fermions, while valence quarks are
overlap fermions. 
As a first application we have studied the spectrum of the Neuberger-Dirac operator, as well as the topological susceptibility, in the background of
dynamical configurations at $a\approx 0.07~\fm$, for three values of the Wilson sea quark mass. The  mixed-regime of ChPT provides predictions for these observables and their sea quark mass dependence in terms of various $N_f=2$ chiral low-energy couplings: at the NLO, they are $\Sigma$, $L_6$ and the combination $L_8+ 2 (L_6+L_7)$. We find that these NLO predictions describe our data well, and the extracted LECs are in good agreement with previous determinations.

This exploratory study obviously needs several important refinements to quantify  systematic errors in a reliable way. Different volumes should be considered to test the expected finite-size scaling.
Also, larger volumes will allow to augment the number of eigenvalues that
can be safely matched to RMT predictions, which in turn will provide a definitive
assessment of the associated systematic uncertainty.
The lattice spacing  is not known very precisely and an accurate determination is under way  by various CLS groups. Obviously other $\beta$ values need to be considered to attempt a continuum extrapolation. Finally the effects of autocorrelations, that we have observed, would need larger statistics to ensure fully
reliable statistical errors.

Our results show that new PQ setups where sea and valence quarks may lie in different chiral regimes ($p$ and $\epsilon$) are tractable (unphysical) regimes from which chiral physics can be extracted. Mixed actions are adequate to treat such regimes, and constitute an interesting approach  for those applications where chiral symmetry plays an important role.

\begin{acknowledgments}

We thank Andreas J\"uttner for providing some Wilson-valence 2-point correlators and the Coordinated Lattice Simulations\footnote{https://twiki.cern.ch/twiki/bin/view/CLS/WebHome} for 
sharing the dynamical Wilson configurations. Our simulations were performed on the IBM MareNostrum at the Barcelona Supercomputing Center, the Tirant installation at the Valencia University,
as well as PC clusters at IFIC, IFT and CERN. We thankfully acknowledge the computer resources and technical support provided by
these institutions. C.P. and P.H. thank the CERN Theory Division for the hospitality while this paper was being finalized. 
F.B. and C.P. acknowledge financial support from the FPU grant AP2005-5201 and
the Ram\'on y Cajal Programme, respectively.
This work was partially supported by the Spanish Ministry for Education and Science projects FPA2006-05807, FPA2007-60323, FPA2008-01732, FPA2009-08785, HA2008-0057 and CSD2007-00042; the Generalitat Valenciana (PROMETEO/2009/116); the Comunidad Aut\'onoma de Madrid (HEPHACOS P-ESP-00346 and HEPHACOS S2009/ESP-1473);
and the European projects FLAVIAnet (MRTN-CT-2006-035482) and STRONGnet (PITN-GA-2009-238353). 
\end{acknowledgments}

%% file: app1.tex
\section{Locality properties of the Neuberger-Dirac operator}
\label{app:locality}

The Neuberger-Dirac operator $D_N$ has been defined in eqs.~\eqref{D_Ndef},~\eqref{Qdef}. While $Q$ is ultralocal, sign$(Q)=Q/(Q^\dagger Q)^{1/2}$ in general couples all space time points. As a consequence, the locality of the Neuberger-Dirac operator is not granted a priori. In \cite{Hernandez:1998et} it was verified that locality is preserved in the quenched case, for values of the lattice spacing around and above the one we are considering now.
We apply the method used there to our specific case.\\

\begin{figure}[!t]
\begin{center}
\includegraphics[width=10.5cm,angle=0]{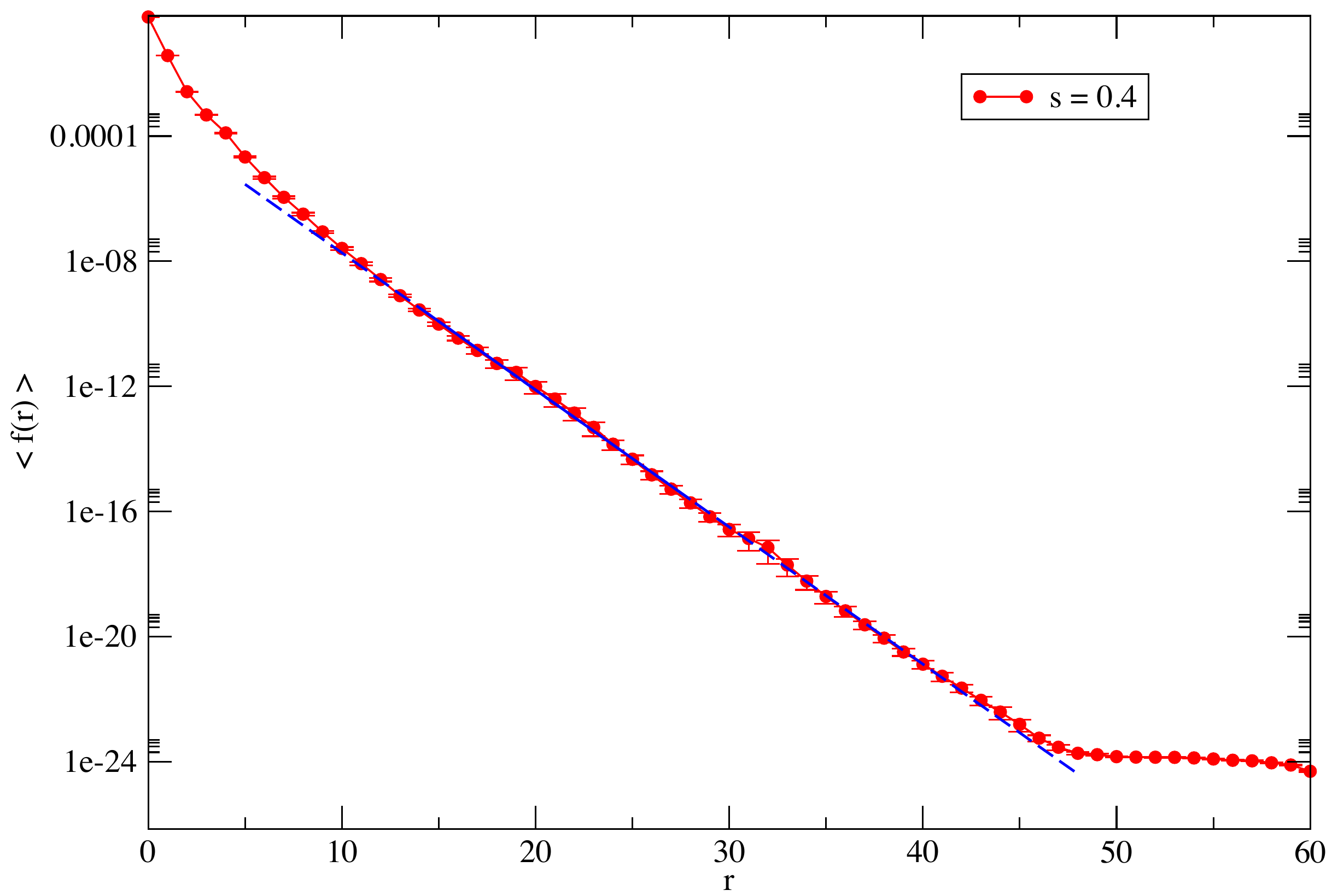}
\includegraphics[width=10.5cm,angle=0]{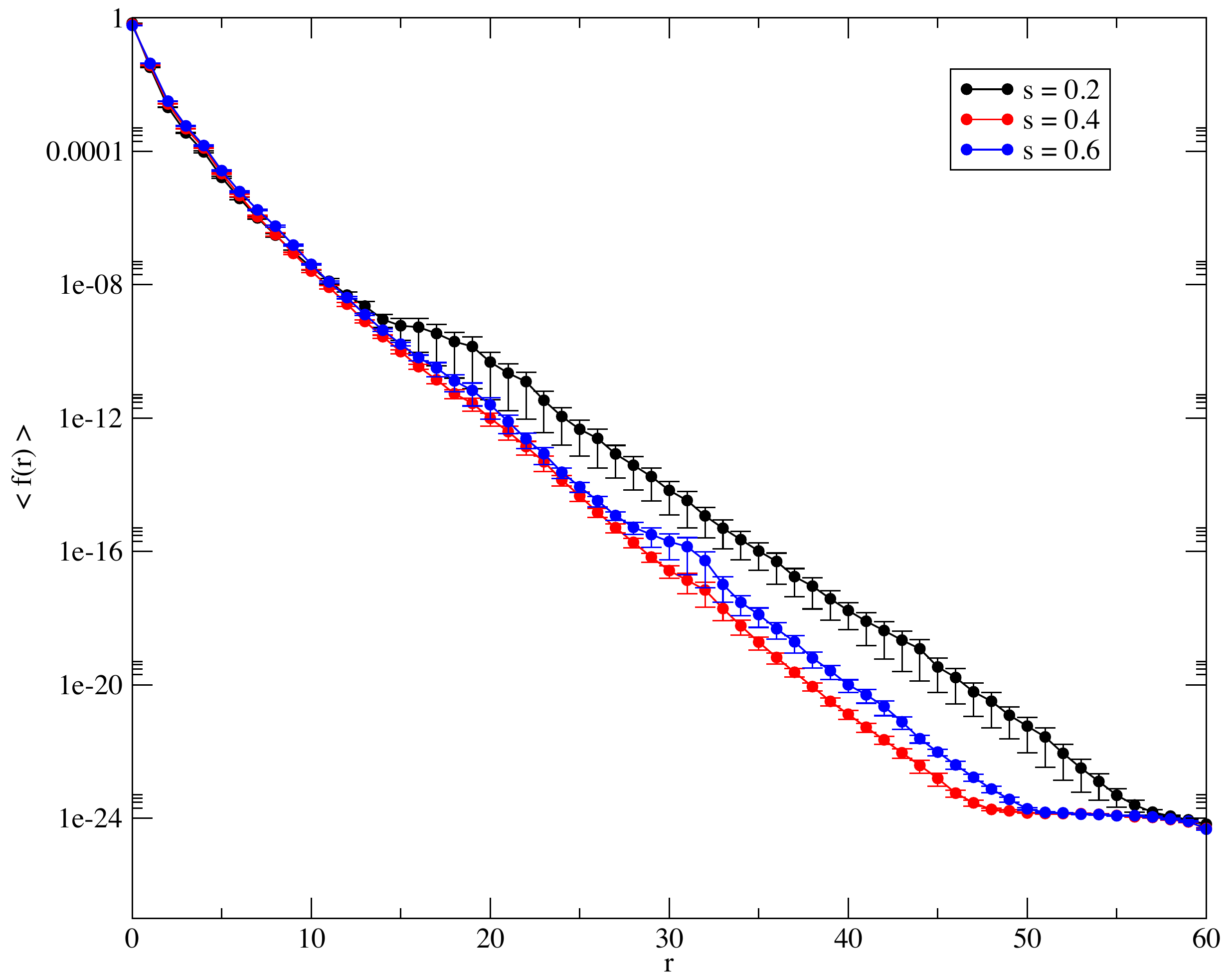}
\end{center}
\vspace{-5mm}
\caption{Upper panel: Result of the fit of $f(r)$ to $Ae^{-Br}$ for $s=0.4$. The blue line is the best fit to the data in the range $14\le r \le28$. Lower panel: Comparison between various values of $s$.}
\label{fig:locality}
\end{figure}

We analyze the effect of the sign operator on a localized field $\eta$:
\begin{equation}
\psi(x)\equiv \mbox{sign(Q)} \eta (x)  \qquad \eta_{ \alpha } (x) = \delta_{xy} \delta_{\alpha 1}
\end{equation}
where $y$ is some point on the lattice and $\alpha $ runs over the colour and the spin indices of the field. We evaluate the function:
\begin{equation}
f(r) \equiv \mbox{max} \{ || \psi (x) || ^2  \mid ||x-y||_1=r  \}\,,
\end{equation}
where $|| \dots ||_1$ denotes the so called ``taxi-driver distance". It is clear that locality is recovered in the continuum limit if $f(r)$ decays exponentially, with rate proportional to the cutoff $1/a$. To check that this indeed happens we fit the lattice data to $A e^{-B r} $ in a range $r_{min} \le r \le r_{max}$. The upper limit has to be set because the inaccuracy with whom we calculate the overlap operator becomes bigger than the value of $f(r)$ at large enough distances.\\ 
The parameters of the simulation were set to calculate reliably $|| \psi (x) || ^2  $ at least down to values of about $10^{-16}$ (corresponding to $r \sim 28$). However, as can be seen in Fig.~\ref{fig:locality}, our implementation of the overlap operator is more precise and the picture of a decaying exponential only breaks down at $r\sim38$ where $f(r) \sim 10^{-22}$.  \\
We have taken in all the cases $r_{min}=14$ and $r_{max}=28$ also to compare with the results in \cite{Hernandez:1998et}. The parameter $s$ in eq.~\eqref{Qdef} can be varied to improve the locality properties of the Neuberger-Dirac operator. We collect the results of our fit for different values of $s$ in Table \ref{table_B}. The quoted error is obtained through jackknife with bin size 1.
Among the values we adopted in our test, the choice $s=0.4$ yields the 
most satisfactory locality properties for this $\beta$ value, as shown in lower panel of Fig.~\ref{fig:locality}. 
This result is in agreement with the previous studies in the quenched case. We will
therefore adopt $s=0.4$ in our study.

\cleardoublepage

\begin{table}[!h]
\begin{center}
\begin{tabular}{|c|c|c|c|}
\hline
s    & $B$ & $10^{4}$A & $\chi^2$\\
\hline
0.2 & $ 0.711 \pm 0.081 $ & $ 0.3 \pm 0.1 $ & 0.70 \\
0.4 & $ 1.001 \pm 0.032 $ & $ 3.4 \pm 1.5 $ & 0.45 \\
0.6 & $ 0.975 \pm 0.031 $ & $ 3.7 \pm 1.2 $ & 0.42 \\
\hline
\end{tabular}
\end{center}
\caption{Results of the fit for dynamical configurations after jackknife resampling. The results for $S=0.2,0.4,0.6$ are based respectively on a set of 13, 21, and 20 configurations, respectively.}
\label{table_B}
\end{table}

%% file: app2.tex
\section{Statistical error analysis}
\label{app:analysis}

\subsection{Autocorrelations}

We have studied the presence of autocorrelations in our observables in various ways:
\begin{itemize}

\item Integrated autocorrelation times have been estimated by using the methods described
in~\cite{Wolff:2003sm} and~\cite{Luscher:2005rx}. In the first case, the summation window
for the normalized autocorrelation function is fixed by setting the parameter $S$
of~\cite{Wolff:2003sm} to $S=2$, while in the second case we stop the summation when
the normalized autocorrelation function is zero within one sigma.

\item The impact of changing the bin size in jackknife resampling of data has been assessed
for all observables.

\item The impact of autocorrelations on statistical errors has been estimated directly with 
the techniques described in~\cite{Wolff:2003sm} for all observables (again with $S=2$).

\end{itemize}
In the case of the D$_6$ lattice, we have studied 
autocorrelations in the D$_{6a}$ ensemble only, as considering it together with the
independent D$_{6b}$ ensemble would result in an underestimation of autocorrelation 
effects.

Our primary observables are the topological charge $\nu$ and expectation values of Dirac
eigenvalues at fixed topology. In order to construct meaningful autocorrelation times
for the latter, we consider their ratio with the RMT prediction for the expectation
value of rescaled eigenvalues, i.e.
$\langle\lambda_k\rangle_{\nu}/\langle\zeta_k\rangle_{{\rm RMT},\nu}=(\Sigma_{\rm eff}V)^{-1}$,
which is $\nu$-independent up to higher orders in the $\epsilon$ chiral expansion.
Results for autocorrelation times are given in Table~\ref{table:tauint}.
While uncertainties on $\tau_{\rm int}$ are remarkably large, due to the fact that
measurements are performed only every 30 HMC trajectories, the values indicate
that results coming from successive configurations are not completely decorrelated,
especially in the case of the D$_6$ lattice.

\begin{table}
\begin{center}
\begin{tabular}{|c|cc|cc|}
\hline
& \multicolumn{2}{|c|}{$\langle\nu\rangle$} & \multicolumn{2}{|c|}{$\langle\nu^2\rangle$} \\[0.1cm]
 & $\tau_{\rm int}^{(1)}$ & $\tau_{\rm int}^{(2)}$ & $\tau_{\rm int}^{(1)}$ & $\tau_{\rm int}^{(2)}$ \\[0.1cm] \hline
& & & & \\[-0.3cm]
D$_4$    & 12.0(6.2) & 8.0(5.2) & 1.47(58) & 0.83(12) \\
D$_5$    & 6.4(3.4) & 3.6(1.7) & 1.7(0.7) & 1.1(0.4) \\
D$_{6a}$ & 4.9(2.5) & 3.8(1.8) & 2.9(1.3) & 2.0(0.8) \\[0.1cm]
\hline
\end{tabular}

\vspace{10mm}

\begin{tabular}{|c|c@{\hspace{3mm}}c|c@{\hspace{3mm}}c|c@{\hspace{3mm}}c|c@{\hspace{3mm}}c|}
\hline
&
\multicolumn{2}{|c|}{$\langle\lambda_1\rangle_{\nu}/\langle\zeta_1\rangle_{{\rm RMT},\nu}$} &
\multicolumn{2}{|c|}{$\langle\lambda_2\rangle_{\nu}/\langle\zeta_2\rangle_{{\rm RMT},\nu}$} &
\multicolumn{2}{|c|}{$\langle\lambda_3\rangle_{\nu}/\langle\zeta_3\rangle_{{\rm RMT},\nu}$} &
\multicolumn{2}{|c|}{$\langle\lambda_4\rangle_{\nu}/\langle\zeta_4\rangle_{{\rm RMT},\nu}$} \\[0.1cm]
 &
 $\tau_{\rm int}^{(1)}$ & $\tau_{\rm int}^{(2)}$ &
 $\tau_{\rm int}^{(1)}$ & $\tau_{\rm int}^{(2)}$ &
 $\tau_{\rm int}^{(1)}$ & $\tau_{\rm int}^{(2)}$ &
 $\tau_{\rm int}^{(1)}$ & $\tau_{\rm int}^{(2)}$ \\[0.1cm] \hline
& & & & & & & & \\[-0.3cm]
D$_4$    & 0.404(94) & $<0.5$   & 1.32(60)    & $<0.5$   & 1.56(77)  & $<0.5$   &  1.22(56) & $<0.5$   \\
D$_5$    & 0.56(15) & 0.89(27) & 1.45(63) & 0.89(27) & 0.93(34) & 0.89(27) & 1.22(54) & 0.89(27) \\
D$_{6a}$ & 2.7(1.3) & 2.3(1.1) & 4.0(2.1) & 3.2(1.5) & 7.0(3.9) & 4.4(2.4) & 7.4(4.0) & 5.2(3.2) \\[0.1cm]
\hline
\end{tabular}
\end{center}
\caption{Estimated autocorrelation times for the topological charge and its square (upper panel),
as well as for the ratios $\langle\lambda_k\rangle_{\nu}/\langle\zeta_k\rangle_{{\rm RMT},\nu}$ up to $k=4$ (lower panel).
Superindices $(1)$ and $(2)$ refer to computations following~\cite{Wolff:2003sm} and~\cite{Luscher:2005rx}, respectively. $\tau_{\rm int}$ is provided in units of gauge configurations, i.e. it should be multiplied times $30$ to convert to number of 
trajectories and times $15$ to convert to MC time.}
\label{table:tauint}
\end{table}

This is further reinforced by the analysis of the dependence of jackknife errors
on the bin size. Again, this exercise is constrained by limited statistics, as
the number of available configurations decreases rapidly with increasing $|\nu|$.
Still, for topologies $|\nu|=0,1,2$ it is possible to have meaningful errors up
to bin sizes of at least 5 configurations on lattices D$_5$ and D$_6$. On lattice
D$_5$, errors for $\langle\lambda_k\rangle_{|\nu|}$ and ratios
$\langle\lambda_k\rangle_{|\nu|}/\langle\lambda_j\rangle_{|\nu|}$ exhibit little or no dependence on the bin size, with errors increasing by at most 20 to 30\%. On the
other hand, on lattice D$_6$ errors consistently increase with the bin size, and,
in cases where there is enough statistics to avoid an early loss of signal, they
tend to saturate around bin sizes of the order of 3--4, at which point they are
between 30\% and 70\% larger than with bin size~1.
Finally, the errors taking into account autocorrelations computed
following~\cite{Wolff:2003sm} are consistent with jackknife errors with bin size~1
on D$_5$, while on D$_6$ they are systematically consistent with those around which
the jackknife bin dependence stabilizes, and in some cases even slightly larger.

Regarding the topological charge, where statistics allows to trace the bin size
dependence up to much larger values, the increase in the error of $\langle\nu\rangle$
and $\langle\nu^2\rangle$ is much more marked than for Dirac eigenvalues.
Typically, jackknife errors stabilize for bin sizes between 5 and 10 for all three 
lattices, at which point they are larger by as much as 60\% with respect to those
for bin size 1.
The analysis \`a la~\cite{Wolff:2003sm} yields comparable errors.

Our conclusion is that there is evidence that autocorrelations affect the topological
charge in all three lattices, while spectral observables are
affected by detectable autocorrelations on D$_6$ only. As a simple recipe
to stay on the safe side when including autocorrelation effect in the errors, we quote 
results for spectral observables from the analysis with jackknife bin size 1 for
lattices D$_4$ and D$_5$ and bin size 3 for lattice D$_6$. In the case of
$\langle\nu\rangle$ and $\langle\nu^2\rangle$ we quote jackknife errors for the
bin sizes at which they stabilize. It has to be stressed that we find no significative 
evidence that autocorrelations depend on the sea quark mass in some systematic way.

\subsection{Systematic errors in the computation of Dirac eigenvalues}

The numerical computation of eigenmodes and eigenvalues of the Hermitian non-negative
operator $D_{\rm N}^\dagger D_{\rm N}$ has been performed with the techniques
described in~\cite{Giusti:2002sm}. The accuracy to which eigenmodes are computed
is bound by an input parameter, which in our case has been set to 1\% in lattices
D$_4$ and D$_5$ and 5\% in lattice D$_6$. On the other hand, for each separate
eigenmode an {\em a posteriori} estimate of the actual error in the computation
of the eigenvalue is produced by the program. Usually, this estimate yields an
error one order of magnitude smaller than the nominal accuracy parameter mentioned
above.

This systematic error should, in principle, be added in quadrature to the statistical
error of $\langle\lambda_k\rangle$ and observables derived thereof. We have estimated
its impact, and found that it is completely negligible with respect to statistical
errors in lattices D$_4$ and D$_5$, both if it is computed using the estimates on
each eigenvalue accuracy and in the much more pessimistic case in which a flat error
associated to the 1\% nominal precision is set. In the case of D$_6$, the uncertainty
coming from numerical error estimates is again negligible, but a flat error set at
5\% yields uncertainties comparable to the statistical ones. However, our experience
shows that the estimates produced by the program are in the right ballpark, and
conclude that setting a flat 5\% uncertainty in D$_6$ observables would be a gross
overestimate of the effect. Hence, we have opted for neglecting this source of
error in our final results.

%% file: figtab.tex
\begin{table}
\begin{center}
\begin{tabular}{|l|c|l|l|l|c|}
\hline
$|\nu|$ &  $k/l$ &  ~~~D$_4$ &  ~~~D$_5$ &  ~~~D$_6$ & qRMT\\
\hline
0 &  2/1  & 3.10(47)  & 2.76(26)&  3.51(36) & 2.70 \\
0 &  3/1 &  5.09(71)  & 5.20(45) & 5.98(72)&  4.46\\
0 &  4/1 & 7.89(96)  &  7.54(65)& 8.7(1.0)&  6.22\\
0 &  3/2 & 1.64(9)     & 1.88(9)  & 1.70(8) &  1.65\\
0 &  4/2 & 2.55(16)  & 2.73(12) & 2.49(15) &  2.30\\
0 &  4/3 & 1.55(7)    &1.45(6)  &1.46(4) &  1.40\\
\hline 
1 &  2/1  & 1.98(10)   &2.08(13) & 2.26(10) & 2.02\\ 
1 &  3/1 & 3.10(16)    & 3.45(23) & 3.56(19) &  3.03\\
1 &  4/1 & 4.30(23)  & 4.80(33) & 4.85(30) & 4.04\\
1 &  3/2 & 1.56(4)  & 1.66(6)  &  1.57(4) & 1.50 \\
1 &  4/2 & 2.17(6)  &  2.31(7)& 2.15(8) & 2.00 \\
1 &  4/3 & 1.39(3)  & 1.39(3) & 1.36(3) & 1.33\\
\hline
2 &  2/1  & 1.98(15)  & 2.00(10) & 2.04(8) &  1.76\\ 
2 &  3/1 & 3.02(21)   &2.99(16) &  3.02(10) &  2.50\\
2 &  4/1 & 3.99(28) & 3.93(20) & 3.94(14) & 3.23\\
2 &  3/2 & 1.53(5)   & 1.49(4) & 1.48(4) & 1.42 \\
2 &  4/2 & 2.02(7)   &  1.96(6) & 1.93(7) &  1.83\\
2 &  4/3 & 1.32(5)  & 1.32(2)  & 1.30(3) & 1.29 \\
\hline
\end{tabular}
\end{center}
\caption{Results for eigenvalue ratios at fixed topology $\langle\lambda_k\rangle^{\nu}/\langle\lambda_l\rangle^{\nu}$ compared with qRMT predictions  $\langle\zeta_k\rangle^{\nu}/\langle\zeta_l\rangle^{\nu}$.} \label{tab_res:ratios1}
\end{table}

\begin{table}
\begin{center}
\begin{tabular}{|l|c|l|l|l|c|}
\hline
$k$ & $|\nu_1|/|\nu_2|$  &  ~~~D$_4$ &  ~~~D$_5$ &  ~~~D$_6$ & qRMT\\
\hline
1 &  1/0  & 2.20(31)  &  1.87(24) &2.12(29) & 1.75\\
2 &  1/0 &  1.41(11)  &   1.41(10)  &1.36(14) & 1.31\\
3 &  1/0 &  1.34(7)  &    1.24(8)& 1.26(10) & 1.19\\
4 &  1/0 &  1.20(4)  &  1.19(6) &1.18(7)  & 1.14 \\
\hline
1  & 2/0  & 2.55(41)      &  2.69(32) &2.76(35)  & 2.45\\
2  & 2/0 & 1.62(14)   & 1.96(14)   & 1.61(16) & 1.59\\
3  & 2/0 & 1.51(11)     & 1.55(10) & 1.40(11) & 1.37\\
4  & 2/0 & 1.29(7)    &  1.40(7) &  1.26(7) & 1.27\\
\hline
1  & 2/1  & 1.16(13)     &1.44(14)  &1.30(11)  &1.40\\
2  & 2/1 & 1.15(8)  & 1.39(9)  &  1.18(8) & 1.22\\
3  & 2/1 & 1.13(8)  &  1.25(7) &  1.12(6) & 1.15\\
4  & 2/1 & 1.07(6)  &  1.18(6)  & 1.07(4) & 1.12\\
\hline
\end{tabular}
\end{center}
\caption{Results for eigenvalue ratios $\langle\lambda_k\rangle^{\nu_1}/\langle\lambda_k\rangle^{\nu_2}$ compared with qRMT predictions  $\langle\zeta_k\rangle^{\nu_1}/\langle\zeta_k\rangle^{\nu_2}$.} \label{tab_res:ratios2}
\end{table}

\begin{table}
\begin{center}
\begin{tabular}{|l| c| l| l| l|}
\hline
$k$ & $|\nu|$ & ~$a^3\Sigma_{\rm eff}$(D$_4$) &  ~$a^3\Sigma_{\rm eff}$(D$_5$) &  ~$a^3\Sigma_{\rm eff}$(D$_6$) \\
\hline
1  &  0   &  0.00135(18) & 0.00121(12) &   0.00089(11)    \\
2  &  0   &  0.00118(8)  & 0.00118(6)& 0.00068(6)\\
3  &  0   &  0.00119(5)  & 0.00104(5)&0.00066(5)\\
4  &  0   &  0.00107(3)  & 0.00100(4)&0.00063(3)\\
\hline
1 & 1 & 0.00108(6) & 0.00113(9) & 0.00074(5)\\
2 & 1 & 0.00110(4) & 0.00110(5)&  0.00066(3)\\
3 & 1 & 0.00105(4) & 0.00099(4)&  0.00063(2)\\
4 & 1 & 0.00101(3) & 0.00095(3)&  0.00061(1) \\
\hline
1 & 2 & 0.00130(13) & 0.00110(7)& 0.00078(2)\\
2 & 2 & 0.00116(7) & 0.00096(5)&  0.00068(3)\\
3 & 2 & 0.00108(7)& 0.00092(4)&  0.00065(2)\\
4 & 2 & 0.00105(5)& 0.00090(3)& 0.00064(2)\\
\hline
\end{tabular}
\end{center}
\caption{The bare effective condensate $\Sigma_{\rm eff}$ extracted from the ratios $\langle \zeta_k \rangle^\nu_{\rm qRMT}/(V \langle\lambda_k\rangle ^\nu_{\rm QCD})$ for $k=1,2,3,4$ and $|\nu|=0,1,2$.} \label{tab:sigmaeff}
\end{table}

\begin{figure}
\begin{center}
\includegraphics[width=12cm]{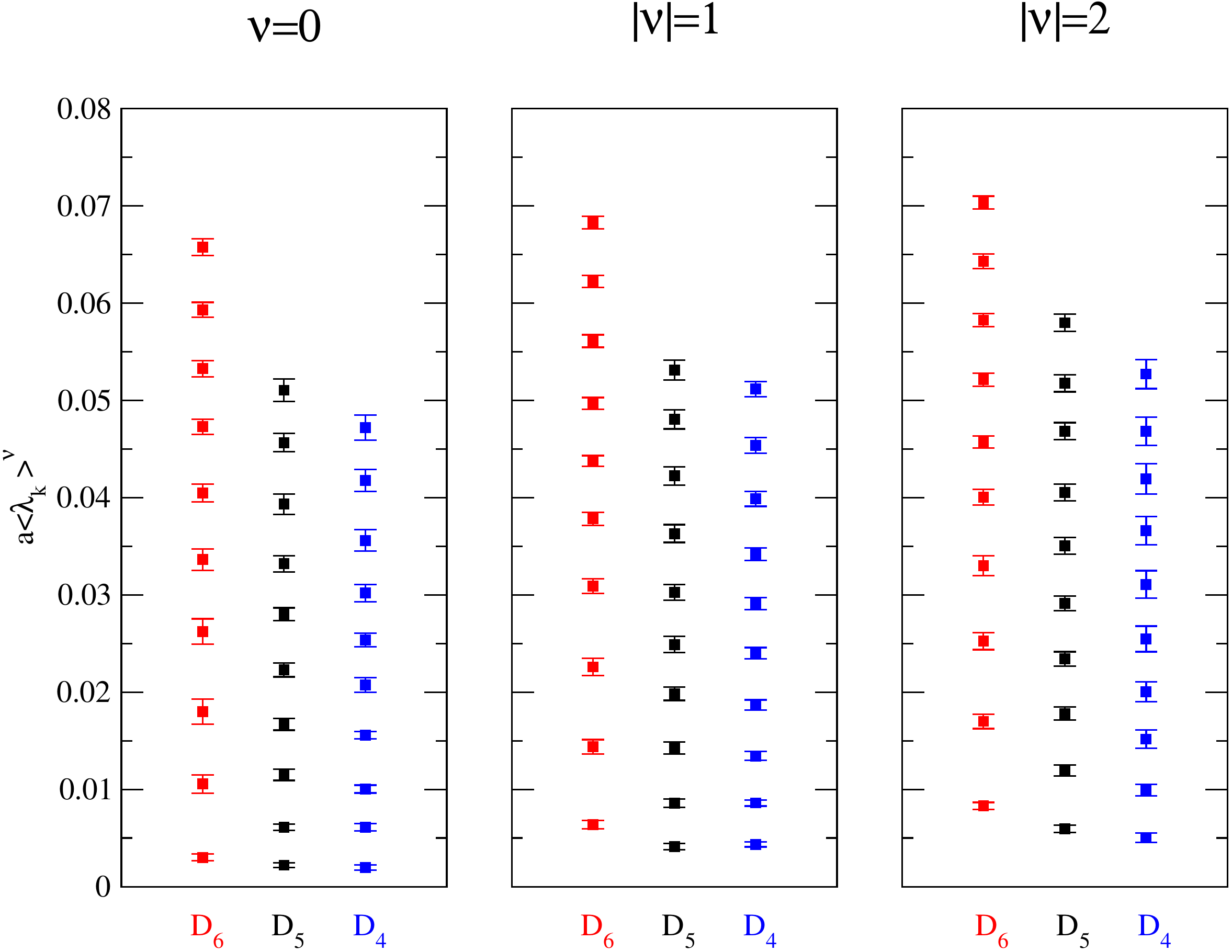}
\caption{The first 10 bare low eigenvalues of the massless Neuberger operator at fixed topology $|\nu|=0,1,2$, for the three lattices D$_4$, D$_5$, D$_6$.} \label{fig:lambda}
\end{center}
\end{figure}

\begin{figure}
\begin{center}
\includegraphics[width=10cm]{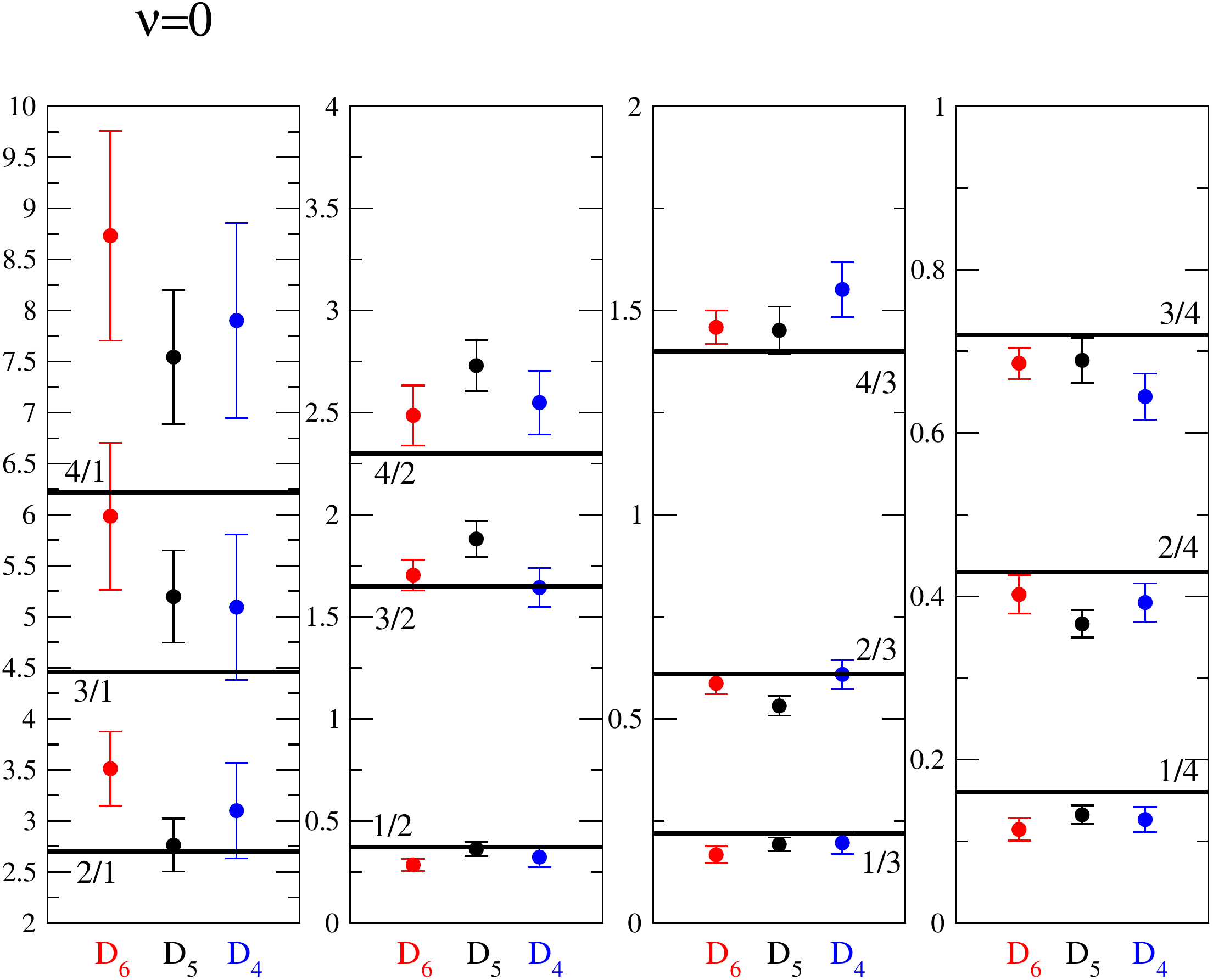}
\caption{Eigenvalue ratios $\langle\lambda_k\rangle^\nu/\langle\lambda_l\rangle^\nu$
at fixed topological charge $\nu=0$. The horizontal lines represent the qRMT prediction.}\label{fig:rat_nu0}
\end{center}
\end{figure}

\begin{figure}
\begin{center}
\includegraphics[width=10cm]{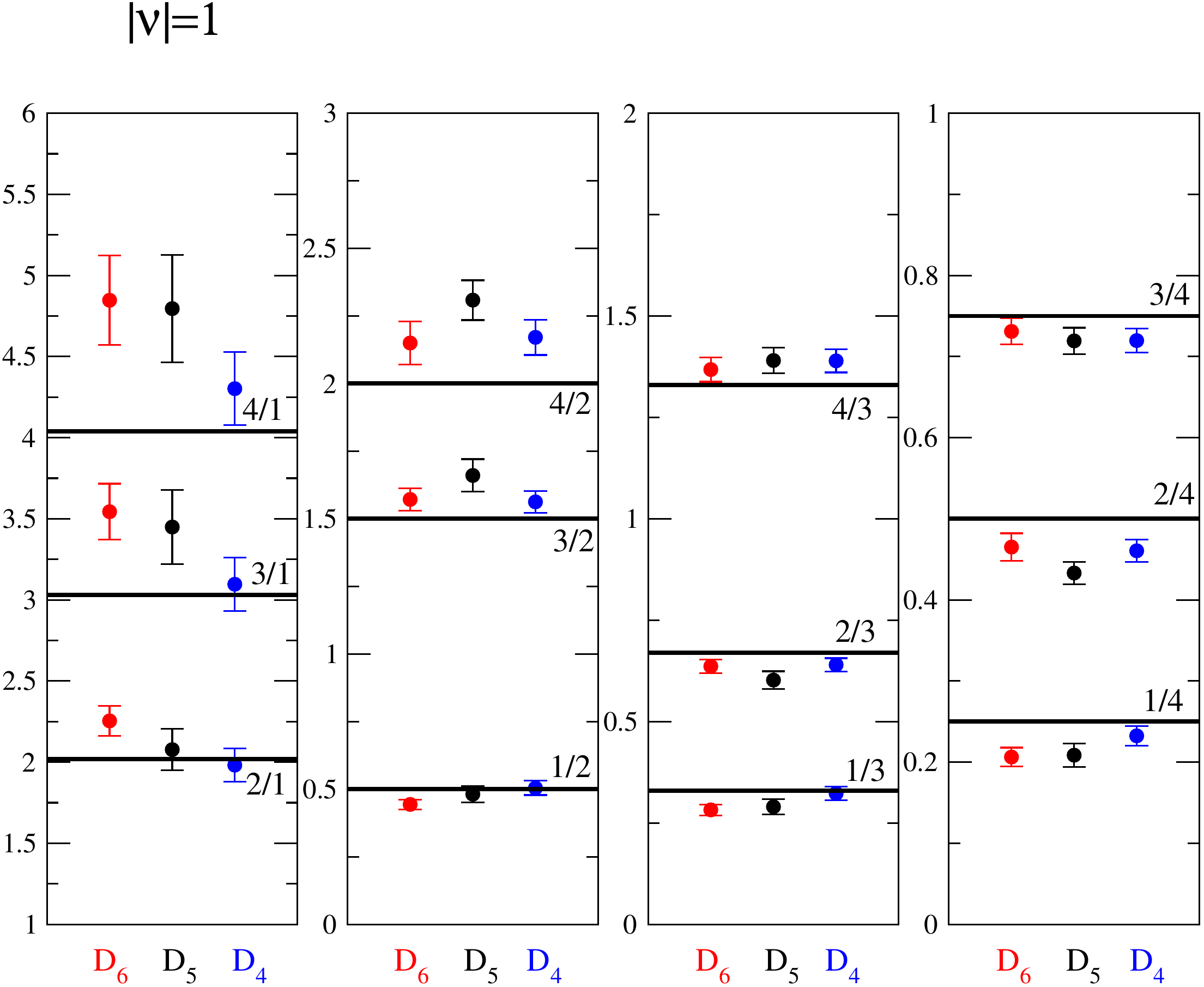}
\caption{Eigenvalue ratios  $\langle\lambda_k\rangle^\nu/\langle\lambda_l\rangle^\nu$ at fixed topological charge $|\nu|=1$. The horizontal lines represent the qRMT prediction.}\label{fig:rat_nu1}
\end{center}
\end{figure}

\begin{figure}
\begin{center}
\includegraphics[width=10cm]{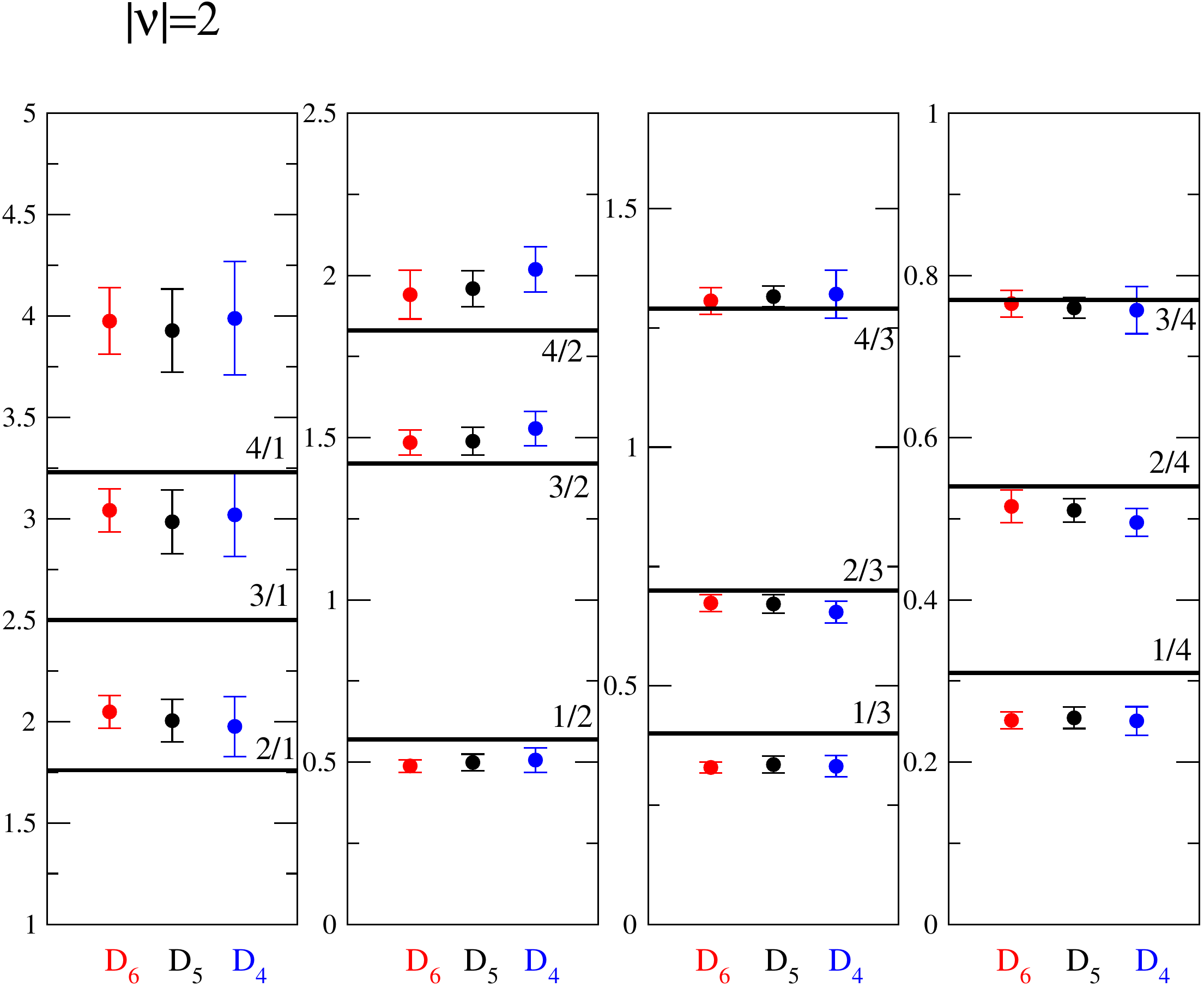}
\caption{Eigenvalue ratios $\langle\lambda_k\rangle^\nu/\langle\lambda_l\rangle^\nu$ at fixed topological charge $|\nu|=2$. The horizontal lines represent the qRMT prediction.}\label{fig:rat_nu2}
\end{center}
\end{figure}

\begin{figure}
\begin{center}
\includegraphics[width=10cm]{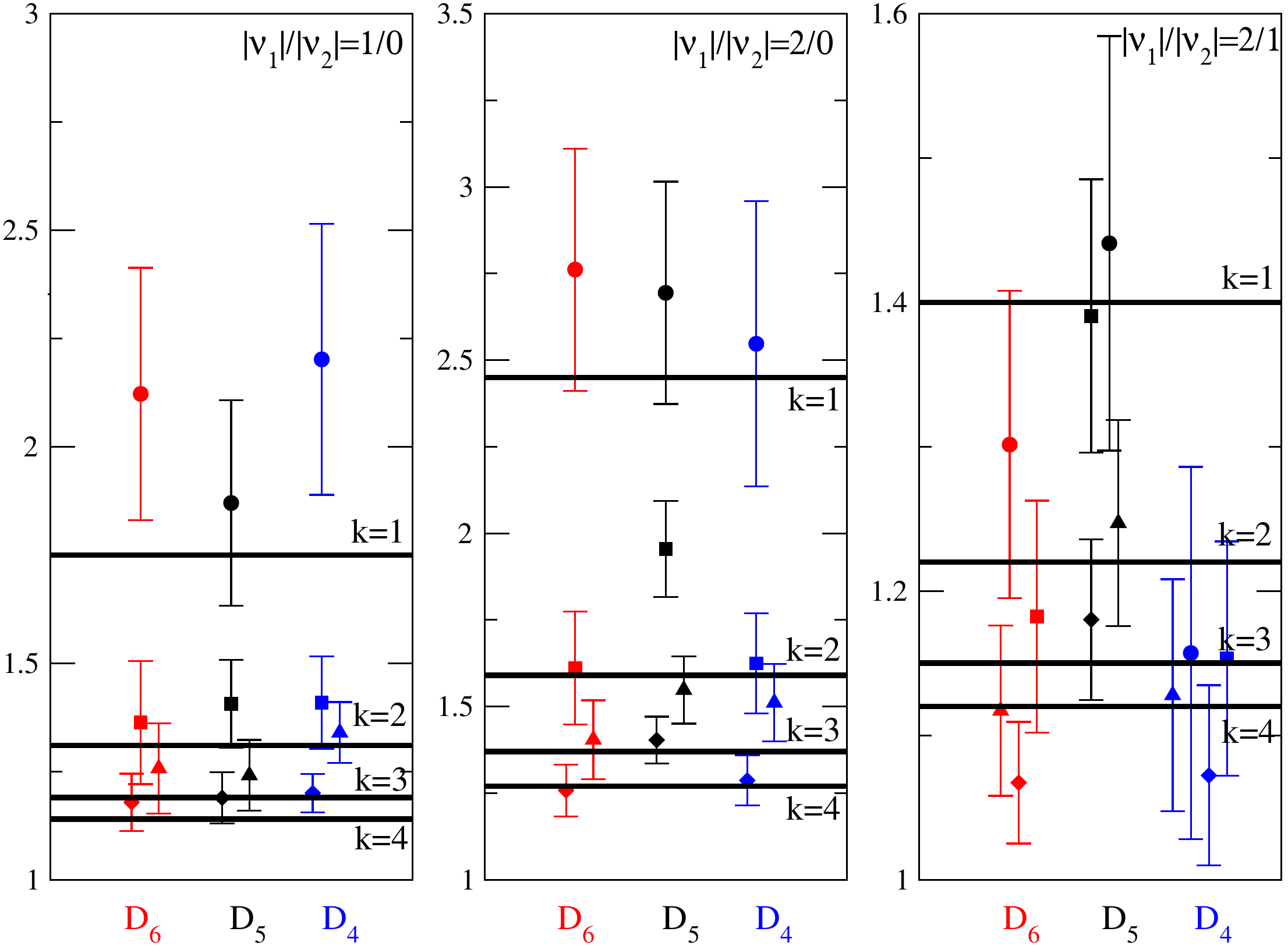}
\caption{Eigenvalue ratios at fixed $k$, for different topological sectors. The circles correspond to $k=1$, the squares to $k=2$, the triangles to $k=3$ and the diamonds to $k=4$. The horizontal lines represent the qRMT prediction.}\label{fig:rat_k}
\end{center}
\end{figure}

\begin{figure}
\begin{center}
\includegraphics[width=10cm]{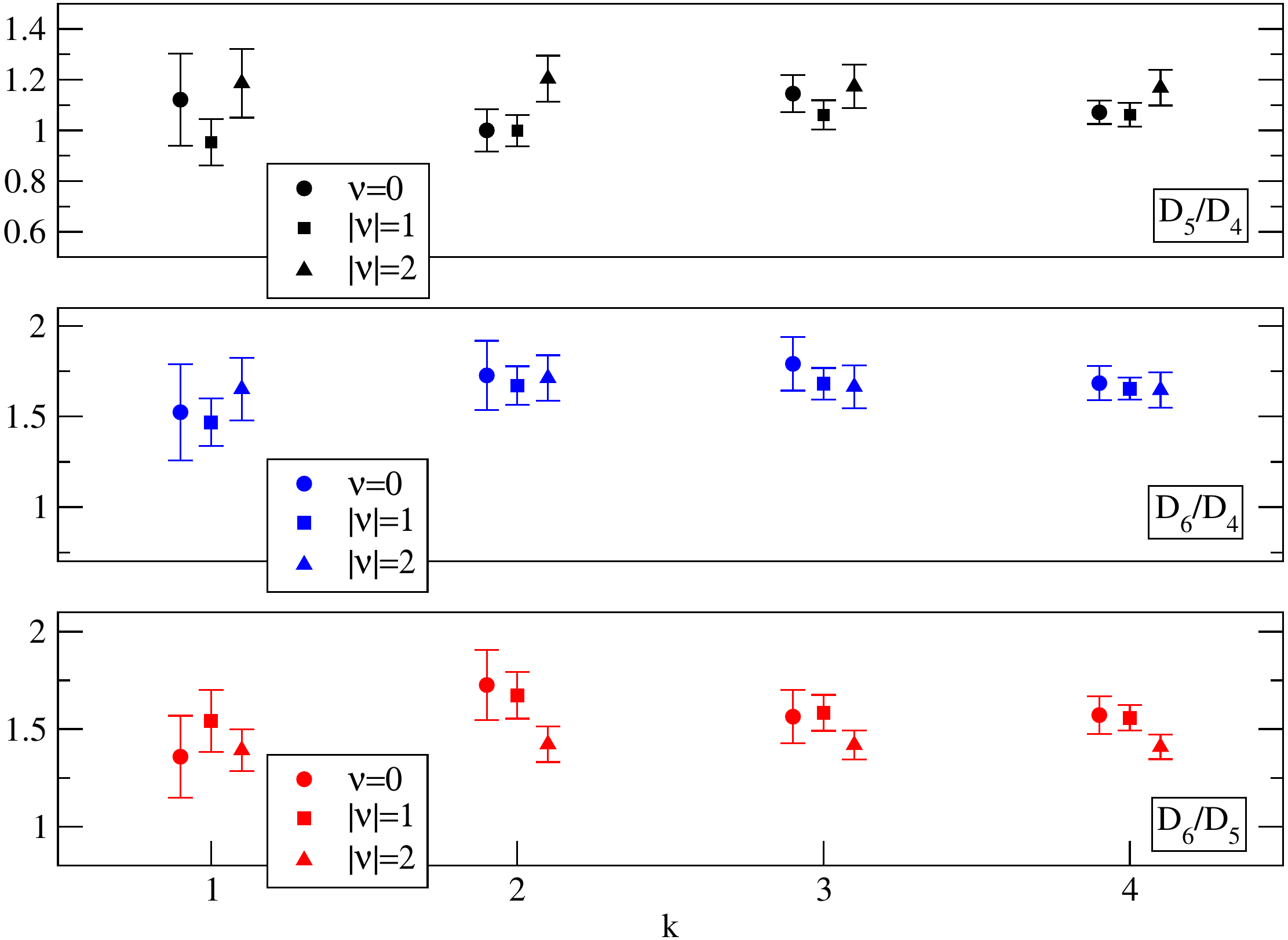}
\caption{Ratios of eigenvalues at different sea quark masses, $\langle\lambda_k\rangle^\nu(M_1) / \langle\lambda_k\rangle^\nu (M_2)$, for $k=1,2,3,4$ and $|\nu|=0,1,2$.}\label{fig:sigmarat}
\end{center}
\end{figure}